# Tracking four-dimensional atomic evolutions of single nanocatalysts throughout the life cycles

**Authors:** Jisheng Xie, Zhiheng Xie, Dijin Jiang, Shiyun Li, Yiheng Dai, Yao Zhang, Mufan Li*, Jihan Zhou*

**Affiliations:**

Beijing National Laboratory for Molecular Sciences, Center for Integrated Spectroscopy, College of Chemistry and Molecular Engineering, Peking University; Beijing, 100871, China.

*Corresponding author. Email: jhzhou@pku.edu.cn; and mufanli@pku.edu.cn

**Abstract:** Structural changes induced by chemical reactions critically determine the catalytic performance and mechanism. However, precise tracking of the three-dimensional (3D) atomic structural evolution of individual bimetallic nanocatalysts remains challenging. Here we develop four-dimensional electrocatalytic atomic-resolution electron tomography, a method for directly tracking 3D atomic rearrangements in identical nanoparticles by electrocatalytic reactions. Using Pd-Pt bimetallic nanoparticles as a model system, we capture the atomic evolution of single nanocatalysts throughout electrocatalytic cycles. We observe two stages of evolutions: surface reconstruction and atom leaching, which are corroborated with the voltage-dependent behaviors probed by in situ electrochemical transmission electron microscopy. We identify chemical short-range order at atomic level and further reveal anisotropic chemical redistributions across different crystallographic orientations. These findings highlight the necessity of incorporating 3D spatiotemporal and chemical evolutions into the rational design of functional nanocatalysts in the future.

**Main Text:** The interplay between distinct metal components of nanocatalysts underlies the structures with their performance in catalysis (*1, 2*). During reactions, catalyst structures evolve through processes like surface reconstruction (*3, 4*), segregation (*5–7*) and atom leaching (*8, 9*). Understanding these atomic-scale structural changes is crucial for elucidating activation and deactivation mechanisms, thereby enabling the rational design of novel catalysts. Transmission electron microscopy (TEM) offers the highest spatial resolution among characterization techniques and has seen decades of improvement in temporal and dimensional resolution. Its observation timescale has advanced from ex situ (*8, 9*) to in situ (*10–12*) and operando (*13–15*) regimes. However, conventional two-dimensional (2D) TEM projections suffer from an intrinsic limitation: the overlapped information along the beam direction, which persistently risks erroneous structural analysis. To resolve this dimensional challenge, methods for three-dimensional (3D) atomic-resolution reconstruction have been progressively developed, including atomic-resolution electron tomography (AET) (*16–19*), discrete tomography based on atom counting (*20–23*), and single-particle reconstruction in graphene liquid cell (*24–26*). These techniques enable direct correlation of a material's 3D atomic structures with its properties (*27–31*). Presently, AET is capable of experimentally resolving the 3D atomic structures of multicomponent nanoparticles (NPs) (*32–34*). In the meantime, while spatiotemporal resolution remains a key pursuit, Zhou et al. have developed four-dimensional AET (4D-AET) to capture early nucleation events at atomic level through continuous heating in vacuum (*35*). Nevertheless, observing the 4D atomic evolution of the same bimetallic NP throughout catalytic reactions has remained elusive, hindered by experimental challenges in tracking the identical nanoparticles at the identical locations and the additional complexity of catalytic reactions. To address this gap, here we develop 4D electrocatalytic AET (4D-ecAET) to directly track atomic structural evolutions of single nanocatalysts triggered by electrocatalytic reactions. Using Pd@Pt NPs, a highly uniform model system widely employed for studying bimetallic interactions (*36–39*), we successfully tracked the atomic evolutions of three sets of Pd@Pt NPs across continuous electrocatalytic cycles. The structural changes were initially dominated by surface reconstruction stage, followed by the stage of atom leaching. After prolonged cycling, we observed orientation-dependent anisotropic evolution in chemical short-range-order parameters (CSROPs), which subsequently led to diverse elemental enrichments on the catalyst near-surface regions. In situ electrochemical TEM observations show voltage-dependent dynamic behaviors for surface reconstruction and atom leaching, which further lead to the diverse structures after continuous cycles. We anticipate that this work will provide a valuable method for tracking 4D structural evolutions of nanocatalysts at single-particle level, which could pave a new way for the correlation of structures with catalytic performance and offer rational guide for designing new corrosion-resistant catalyst architectures.

## Achieving four-dimensional electrocatalytic atomic-resolution electron tomography

Pd@Pt NPs were synthesized as 6-7 nm cuboctahedrons with core-shell morphology and face-centered cubic symmetry (fig. S1, A and B), supported on carbon nanotubes (CNTs), and characterized by aberration-corrected annular dark-field scanning transmission electron microscopy (ADF-STEM) and energy-dispersive X-ray spectroscopy (EDS) (see fig. S1, C to H and materials and methods). Catalyst inks (Pd@Pt/CNT) were deposited on gold mesh grids (table S1); and the grids were then served as both TEM specimens for tilt series acquisition and working electrodes for electrocatalytic durability testing (Fig. 1A). Then we carried out the demonstration experiments to check the feasibility of 4D-ecAET (see materials and methods). First, using electrocatalytic ethanol oxidation reaction (EOR) as a probe reaction (*40–42*), we have confirmed the reaction activity on the home-made working electrodes with Pd@Pt/CNT using gold mesh grid; and no activity was observed on the control working electrodes with catalyst-free gold mesh grid (fig. S2). Second, the same Pd@Pt NPs at identical locations were tracked before and after sequential durability cycles, with tilt series acquired at each time interval. In each run of data acquisition, low-magnification images around the targeted NPs were acquired to record the surrounding information of NPs and CNT morphology (fig. S3), enabling precise relocation of the targeted NP even after 2.5K cycles (fig. S3, F and H). This enabled successful 4D-ecAET observation of structural changes in Pd@Pt NPs throughout electrocatalysis.

Three sets of tomographic tilt series for 4D-ecAET (particles 1-3, seven datasets in total) were acquired in ADF modes of an aberration-corrected STEM (see figs. S4 to S12 and materials and methods) under lower electron doses (table S2) compared to the previous study (*35*). The name for each particle at different stages of cycles is Pd@Pt$_{m\_n}$ (m = 1, 2, 3 and n = 0, 0.1K, 0.2K, 0.5K, 1.5K and 2.5K), where m means the number of particle series and n means the number of electrocatalytic cycles in total. Corresponding electrocatalytic curves and peak currents show the loss of activity in each batch (fig. S13). Representative images at similar zone axis for particles 1-3 show the structural changes become severe as the cycles of reactions increased (Fig. 1, B and C, fig. S14). By implementing image preprocessing, 3D reconstruction, atom tracing, and classification procedures, we obtained the final experimental 3D atomic model for each particle at different cycles (see Fig. 1, E to G, fig. S15 and materials and methods). Particles before reaction displayed the cuboctahedral morphology with element-mixing surfaces (Fig. 1E, fig. S14C and fig. S15D), consistent with our recent observations (*28, 43, 44*).

By comparing the 3D experimental models before and after reactions, two stages of structural evolutions were observed throughout the cycles of reactions: surface reconstruction (≤0.5K cycles) and atom leaching (>500 cycles): (1) for the stage of surface reconstruction, we first compared the 3D structure of particle-1 before reaction with the one after 0.5K cycles (Fig. 1, E and F). surface reconstruction happened and the corner part got rounded near the {111} facet (Fig. 1F). To confirm the limited structural changes at the early stage, we compared transitions <0.5K cycles: particle-2 after 100 cycles showed minor surface reconstruction (fig. S15, A and B), particle-3 after 200 cycles in 2D images exhibited analogous rounding in shape like that in particle-1 after 500 cycles (fig. S14, C and D). (2) For the stage of atom leaching, after 1.5K cycles, the size of particle-1 become smaller, indicating severe atom leaching (Fig. 1, F and G). To confirm the severe structural changes after longer cycles, we compared transitions >1.5K cycles: particle-3 after 2.5K cycles displayed atom leaching and severe Pd segregation (fig. S15D). We repeated identical experiments to get more images in 2D for six additional NPs (fig. S16). All NPs universally underwent changes in shapes and atom leaching after continuous cycling. Although the exact cycles for structural transformation varied among particles, NPs followed the order of "surface reconstruction - atom leaching" evolutions. All these results show surface reconstruction dominates initially and atom leaching happens afterwards. These observations suggest vacancy-mediated atomic diffusion mechanisms plays critical role in the EOR reaction, where surface vacancy formation is energetically favored over bulk vacancy generation (*45*).

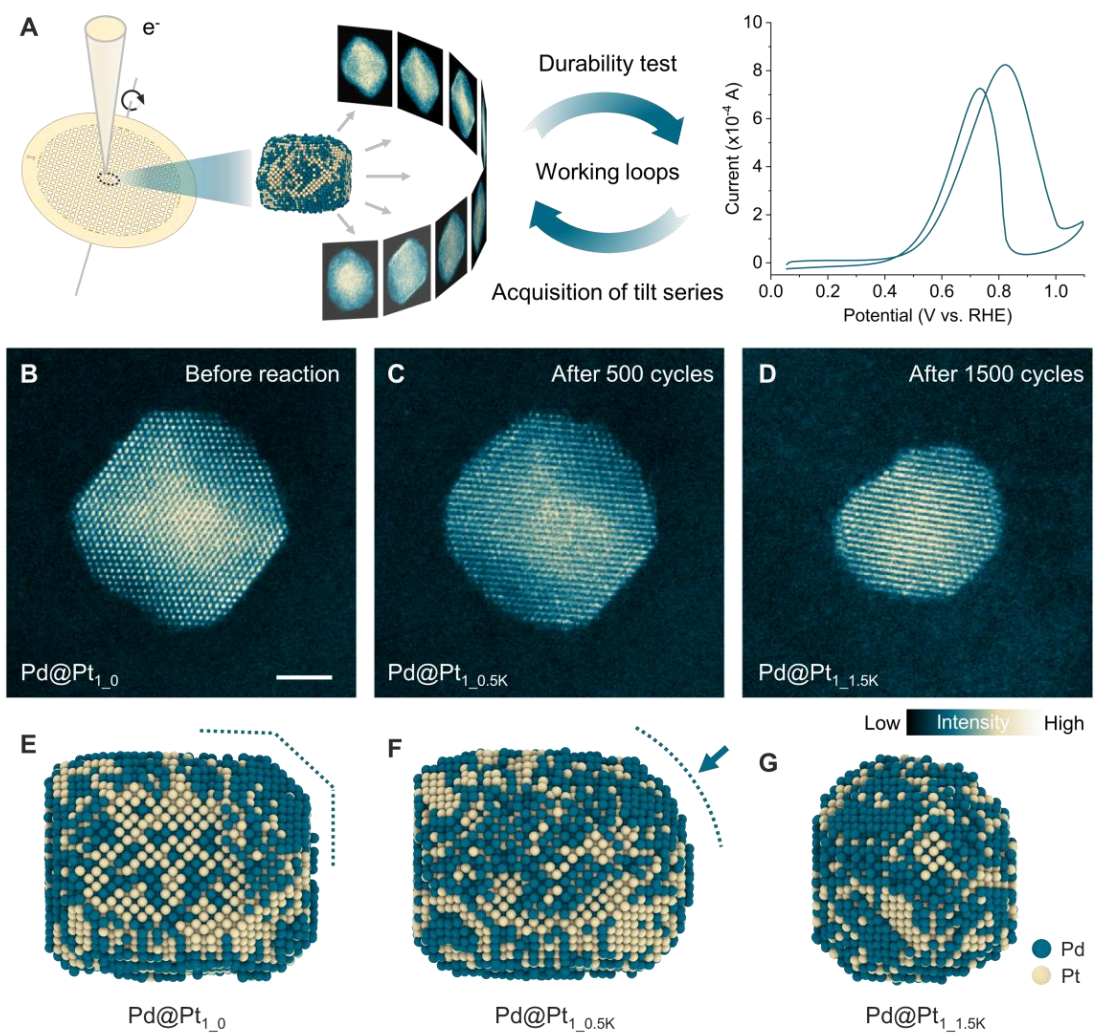

**Fig. 1. Four-dimensional electrocatalytic atomic-resolution electron tomography.** (**A**) Schematic of integrated work loops combining atomic-resolution electron tomography (left) with electrocatalytic testing (right) to track 3D structural evolutions of nanocatalysts throughout catalytic cycles. Gold mesh grids function dually as specimen supports for tilt-series acquisition and working electrodes for electrocatalytic durability tests, with both processes iterated during experiments. (**B** to **D**) ADF-STEM images from tomographic tilt-series of particle-1 (B) before reaction, (C) after 500 cycles and (D) after 1,500 cycles, respectively. The tilting angles for the images are (A) -37°, (B) -31° and (C) -37°, respectively. Images were registered from three frames and normalized. The colormap for (B) to (D) is placed below (D). Scale bars are 2 nm for (B) to (D). (**E** and **G**) Experimentally derived 3D atomic renderings for (E) Pd@Pt$_{1\_0}$, (F) Pd@Pt$_{1\_0.5K}$ and (G) Pd@Pt$_{1\_1.5K}$. Pd and Pt atoms are depicted as green and yellow spheres respectively. The dashed lines show the evolution of particle shapes, which gets rounded after 0.5K cycles. Dashed lines and the arrow highlight the places of rounding in shape for particle-1 after 500 cycles.

### Tracking atomic evolutions in positions and chemical species

General structural analysis, including atom numbers, specific surface area, surface coordination number (CN) and facet distribution, was performed based on all 3D atomic models (see materials and methods). For datasets ≤0.5K cycles, while atom numbers and specific surface area remained nearly identical for Pd@Pt$_{1\_0}$ Pd@Pt$_{1\_0.5K}$ (fig. S17A), surface CN fluctuates and the proportion of high-index facets ("other") increases (fig. S17, B and C). This indicates that the rounded shape in Pd@Pt$_{1\_0.5K}$ is mostly attributed to surface reconstruction rather than etching at this stage. Similar but milder surface reconstruction features appeared in particle-2 after 0.1K cycles (fig. S17, D to F). For particles experienced more than 0.5K cycles, 53% of atoms were leached from particle-1 after 1.5K cycles and 57% of atoms from particle-3 after 2.5K cycles (table S2), leading to deceased surface area, increased specific surface area and dramatic shifts in surface CN and facet distribution (fig. S17).

To quantify the 3D atomic evolutions in structural changes, evolutions in positions and chemical species were tracked. First, atom moves would create disappeared and new positions between two datasets. We tracked the identical atomic positions in each particle across cycles, named as common positions (positions that exist in both of two datasets, see materials and methods). During the stage of surface reconstruction, for particle-1, 96% of atoms were paired as common positions between Pd@Pt$_{1\_0}$ and Pd@Pt$_{1\_0.5K}$ (table S2). The unpaired positions were classified as disappeared atoms (present only in Pd@Pt$_{1\_0}$, Fig. 2A, upper panel) or new atoms (only in Pd@Pt$_{1\_0.5K}$, Fig. 2A, central panel). The unpaired atoms are predominantly Pd atoms (Fig. 2B) and located within 2-Å distance to the surface (Fig. 2C), indicating high Pd mobility during surface reconstruction (*37*). The peak positions of CN for unpaired atoms shift from 6 towards 8 (from disappeared atoms to new atoms) after 0.5K cycles (Fig. 2D), revealing the major direction of surface reconstruction from edges to {100} sites. Second, migrated atoms might exchange their positions, and this would lead to different chemical species on the common positions. To check this phenomenon, we calculated the consistency of chemical species for common positions, 89% of atoms are consistent in particle-1 and recognized as a core (Fig. 2A, lower panel). These inconsistent atoms (Fig. 2E) flip from one to another position during the reaction, migrating mostly within 5-Å distance to the surface (~1-2 atomic layers, Fig. 2F). Particle-2 shows similar behaviors during the stage of surface reconstruction (fig. S18), 92% atoms are consistent between 0 and 0.1K cycles due to the shorter cycles (table S2). While our previous reports demonstrated that AET achieves approximately 95-97% structural consistency when repeatedly measuring the same particles twice (95% for FePt NPs (*35*) and 97% for Pd@Pt NPs (*43*)), this indicates that structural changes after 100 cycles for particle-2 in this work are very minimal and approaching the limit of AET.

During the stage of atom leaching, for particle-1 (0.5K-1.5K cycles), severe atom leaching happened (Fig. 2G), the number of paired atoms reduced (Fig. 2H), and only 49% of atoms in Pd@Pt$_{1\_0.5K}$ were paired as common positions (table S2). Disappeared atoms are mostly Pd atoms (Fig. 2H). Atom leaching happens in deeper regions of 20-Å distance to surface (Fig. 2I); and particle-1 lost more than 4,000 Pd atoms with CN of 12 (fully-coordinated atoms, Fig. 2J), showing continuous exposure of inner Pd atoms to the electrolyte and subsequent oxidative etching (*46, 47*). Only 36.4% of atoms in Pd@Pt$_{1\_0.5K}$ remain consistent with Pd@Pt$_{1\_1.5K}$ as a core (Fig. 2G, lower panel). Migrated inconsistent atoms (Fig. 2K) were primarily Pd, confirming Pt's lower mobility, and the migration depths also reached 20 Å to surface (Fig. 2L). After thousands of electrocatalytic cycles, sustained near-surface reconstruction ultimately exposes inner Pd, accelerating atom leaching in the following cycles.

To investigate the dynamic behaviors during reaction, in situ electrochemical TEM was employed to track the 2D structural evolutions of individual NPs during the reaction (see materials and methods). Prior to imaging, catalysts were activated via cyclic voltammetry (CV) from 0.28

V vs. reversible hydrogen electrode (V$_{RHE}$) to 1.08 V$_{RHE}$ in the EOR condition. We monitored the morphology changes of NPs during CV (fig. S19 and movie S1). As potential increased, the NP got rounded (representative TEM images in fig. S20) after 0.5 V$_{RHE}$, indicating surface reconstruction observed in AET results origins from the atomic diffusion from this oxidation potential (fig. S14). Given Pd's lower oxidation potential versus Pt (*48*), it can be inferred that Pd would own a higher mobility under higher oxidation potentials, leading to greater exchanges in Pd positions (Fig. 2B). To probe dynamics at high oxidation potential, chronoamperometry (CA) was applied to a Pd@Pt NP at 0.68 V$_{RHE}$ (fig. S21 and movie S2). A fluctuating liquid-like cluster with a certain extent of contrast (lighter than the NP) entered the field of view (0.8 s, fig. S22), adhered to the NP edge (0.8-1.2 s), grew up into a sphere-like shape (1.2-1.6 s), and diffused locally along the NP corner. After 5.2 s, the cluster disappeared in the electrolyte (fig. S21), suggesting atomic etching and dissolution into the electrolytes as ionic form. Similar behavior was observed in another NP under CA at 0.68 V$_{RHE}$ (figs. S23 and S24 and movie S3). To check these structural changes were not electron-beam induced, we performed the control experiments using identical imaging conditions. Two NPs were imaged with no bias, and under CA test with a lower potential of 0.28 V$_{RHE}$, respectively (movies S4 and S5). No structural changes like rounding shape and atom leaching were observed in these two NPs (fig. S25), the generation of rounded shape and leaching clusters are intrinsic structural changes under electrochemical conditions at higher oxidation potentials and independent of electron beam irradiation. These results under reaction conditions indicate that continuous exposure at higher oxidation potential would trigger the atom leaching process, and it explains prolonged cycling promotes atom leaching, reducing particle size and altering near-surface chemical environments.

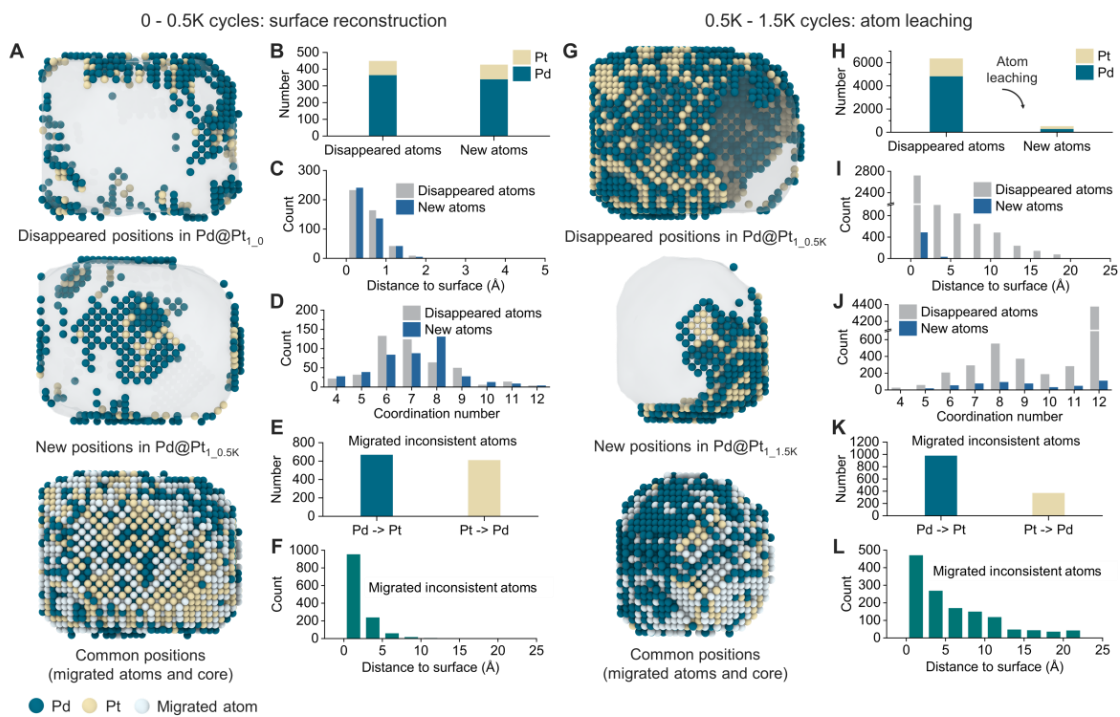

**Fig. 2. 3D structural evolution in particle-1.** (**A**) 3D renderings before and after 500 cycles: disappeared positions (upper panel, positions in Pd@Pt$_{1\_0}$ only), new positions (central panel, only in Pd@Pt$_{1\_0.5K}$), and common positions (lower panel, common positions during the reaction).

Green, yellow, and white spheres are denoted as Pd, Pt, and migrated inconsistent atoms at the common positions, respectively. (**B** to **D**) Basic structural analysis of unpaired atoms before and after 500 cycles: (B) numbers of atoms, (C) distribution of distances to surface and (D) distribution of coordination numbers. In the calculation of distance to surface: distances of disappeared atoms are calculated by Pd@Pt$_{1\_0}$ surface, and distances of new atoms are calculated by Pd@Pt$_{1\_0.5K}$ surface. (**E** and **F**) Migrated inconsistent atoms at common positions before and after 500 cycles: (E) numbers of atoms and (F) distribution of distances to surface (calculated by Pd@Pt$_{1\_0}$ surface). (**G** to **L**) Similar analysis between 500 and 1,500 cycles: (G) 3D renderings, (H) numbers of atoms, (I) distribution of distances to surface (disappeared: calculated by Pd@Pt$_{1\_0.5K}$ surface; new: calculated by Pd@Pt$_{1\_1.5K}$ surface), (J) distributions of coordination numbers, (K) migrated atom numbers at common positions and (L) distributions of their distances to surface (calculated by Pd@Pt$_{1\_0.5K}$ surface).

**Anisotropic evolutions in the chemical short-range-order parameter**

To further investigate the chemical evolution in near-surface regions, we tracked the elemental redistributions across continuous catalytic cycles. We quantified the spatial heterogeneity in near-surface regions of particle-1 (see materials and methods). First, we calculated the Pd concentrations at different depths (Fig. 3A). Similar curves were observed in particle-1 before and after 0.5K cycles. After 1.5K cycles, Pd concentrations decreases in 2$^{nd}$ and 3$^{rd}$ shells. Next, we characterized particle-1's local chemical heterogeneity by analyzing heterogeneous centers (defined as central atoms coordinated by both Pd and Pt neighbors, Fig. 3B) and local Pd concentrations ($c_{Pd}$) within coordination shells of Pt centers (Fig. 3C). Heterogeneous centers and the peak positions in $c_{Pd}$ distributions for Pd@Pt$_{1\_0}$ and Pd@Pt$_{1\_0.5K}$ are similar, while the counts decrease in Pd@Pt$_{1\_1.5K}$ due to atom leaching (Fig. 3B). This interestingly shows that features of local chemical heterogeneity could be preserved in total after long cycles. To investigate whether the elemental redistributions are isotropic in different regions of surfaces, CSROPs were calculated to map out the Pd/Pt enrichments in particle-1 (Fig. 3D). Anisotropic evolution was observed: in the front views of particle-1, we found CSROPs increases in specific regions, transitioning from Pt-rich (yellow) to Pd-rich (green) after 1.5K cycles; in the rear view of particle-1, we found disappearance of Pd-rich regions. To resolve the crystalline orientation dependence, we further divided particle-1 into 14 regions (with six {100} orientations and eight {111} orientations) based on its cuboctahedral symmetry (see materials and methods). The delta values (differences between datasets) were then calculated based on the orientation-classified averages of CSROP values (Fig. 3E, table S3). In particle-1, {111} regions exhibited significant CSROP changes, while {100} regions show less changes in average.

Similar analysis was also carried out for particle-2 (fig. S26) and particle-3 (fig. S27): For particle-2 after 100 cycles, where surface reconstruction dominates, the increased surface Pd concentration of Pd@Pt$_{2\_0.1K}$ also indicates surface reconstruction yields a Pd-abundant surface (fig. S26A). Local chemical heterogeneity (fig. S26, B and C) and CSROPs (fig. S26, D and E) during 0-100 cycles exhibit minimal changes and similar trends. Notably, from all crystalline orientations of particle-2, delta values of averaged CSROPs between Pd@Pt$_{2\_0.1K}$ and Pd@Pt$_{2\_0}$ are below ±0.2 (fig. S28B and table S3), consistent with our observation of a lower-degree of surface reconstruction. For particle-3, severe atom leaching after 2.5K cycles elevated surface-shell Pd concentration from 53% to 86% (fig. S27A). The numbers of heterogeneous Pd centers increased while those of Pt centers decreased at the surface (fig. S27B). The Pd-enriched surface consequently shifts $c_{Pd}$ to positive values (fig. S27C), reflecting more Pd-abundant coordination shells in the overall states. For CSROP evolutions in particle-3 (fig. S27, D and E), more orientations show significant changes in delta CSROPs (fig. S28C and table S3), including both {111} and {100} orientations. This is different from the observation in particle-1 after 1.5K cycles, probably due to the evolutionary differences among NPs. Notably, extended cycling after 2.5K cycles induces more Pt-rich regions in {100} regions of CSROPs (fig. S27E). While the surface atoms are mostly covered by Pd atoms (fig. S15D and S27A), i.e. dominated by Pd segregations, the Pt-rich regions in CSROPs indicate Pt accumulating at subsurface structures. This Pd-rich surface and Pt-rich subsurface structures can be attributed to the lower surface energy of Pd (2.05 J m$^{-2}$) compared to that of Pt (2.48 J m$^{-2}$) (*49*), which drives Pt atoms inward and finally leads to the reduced heterogeneous centers and elevated $c_{Pd}$ values (fig. S27 B and C). To cross-validate the exist of elemental redistribution, we imaged four additional NPs in 2D after 2K cycles without prior beam irradiation (see fig. S29, A to D and materials and methods). The core-shell features are lost among all NPs, which are in irregular and sub-5-nm morphologies as a result of atom leaching. EDS mapping (fig. S29, E to H) reveals an asymmetric Pt distribution within the NP (fig. S29G), which shows elemental redistribution after extended cycling.

These observations demonstrate orientation-dependent anisotropic chemical evolution during electrocatalysis, drastically increasing the surface structural complexity, rationalizing the inconsistent correlation between traditionally proposed structures and activities. After tracking the atomic evolutions among three NPs in 4D, we found two main stages during electrocatalysis which are responsible for the deactivation of Pd@Pt nanocatalyst: (1) the activity changes around 0-500 cycles are highly linked with surface reconstruction; (2) beyond 500 cycles, atom leaching occurs and finally generates anisotropic Pd-rich surfaces with Pt-rich subsurfaces, one of the structural features at which point >90% of activity (fig. S13) was lost (almost every NP has been deactivated).

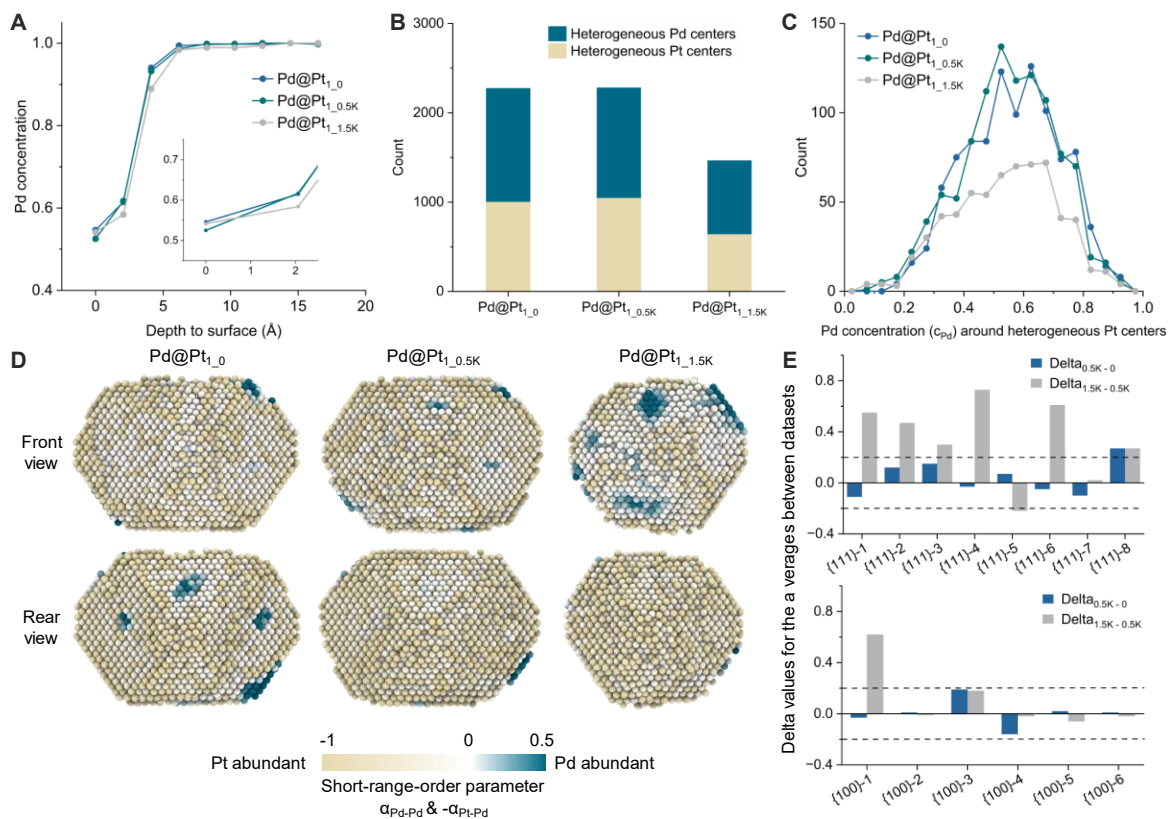

**Fig. 3. Evolution of chemical heterogeneity and short-range-order parameter for particle-1.** (**A**) Shell-by-shell Pd concentrations throughout the evolutions. (**B**) Number of heterogeneous centers, where a center is defined as one atom coordinated by both Pd and Pt neighbors. (**C**) Overall distribution of local Pd concentrations ($c_{Pd}$) around heterogeneous Pt centers. $c_{Pd}$ (x axis) represents Pd/(Pd+Pt) atomic fractions in coordination shells, indicating local Pd occurrence probabilities around central atoms. (**D**) Spatial renderings of CSROPs in front and rear views across cycles. Green represents the Pd-abundant regions, yellow represents the Pt-abundant regions and white represents the regions equal to the averaged Pd concentration in the system. Values were normed to 0.5 or -1, respectively. (**E**) Delta values for the orientation-classified averages of CSROPs. Surface atoms were classified into eight {111} orientations (upper panel) and six {100} orientations (lower panel), with values averaged within each region (figure S28). Then the delta values of $Delta_{0.5K-0}$ and $Delta_{1.5K-0.5K}$ are obtained by calculating the differences of averages between Pd@Pt$_{1\_0.5K}$ and Pd@Pt$_{1\_0}$, and between Pd@Pt$_{1\_1.5K}$ and Pd@Pt$_{1\_0.5K}$, respectively.

**Summary and conclusions**

In summary, we have developed a 4D imaging method for bimetallic NPs throughout electrocatalytic reactions (4D-ecAET) by combing AET with electrocatalytic testing procedures, achieving the tracking of 4D atomic evolutions in single Pd@Pt nanocatalysts throughout their life cycles. Our work reveals a degradation pathway in Pd@Pt nanocatalysts: surface reconstruction initiates via a majority of Pd migration at a distance of 5 Å to the surface, followed by atom leaching penetrating to a distance of 20 Å to the surface. In situ electrochemical TEM confirms the potential-dependent atomic mobility. Crucially, we quantified the anisotropic chemical evolutions in CSROPs, and identified {111} orientations as preferential sites for Pd enrichment during extended cycling. Pd segregation to the surface and Pt accumulation at subsurface are probed as one of the chemical features in the death stage of catalysts. 4D-ecAET establishes a general platform to understand 4D atomic evolutions with functional degradation of energy materials, offering a new characterization tool that combined both temporal- and spatial-scale advantages for the development of next-generation catalysts.

**References**

1. C. Xie, Z. Niu, D. Kim, M. Li, P. Yang, Surface and Interface Control in Nanoparticle Catalysis. *Chem. Rev.* **120**, 1184–1249 (2020). doi: 10.1021/acs.chemrev.9b00220.
2. C. Vogt, B. M. Weckhuysen, The concept of active site in heterogeneous catalysis. *Nat. Rev. Chem.* **6**, 89–111 (2022). doi: 10.1038/s41570-021-00340-y.
3. M. Chi, C. Wang, Y. Lei, G. Wang, D. Li, K. L. More, A. Lupini, L. F. Allard, N. M. Markovic, V. R. Stamenkovic, Surface faceting and elemental diffusion behaviour at atomic scale for alloy nanoparticles during in situ annealing. *Nat. Commun.* **6**, 8925 (2015). doi: 10.1038/ncomms9925.
4. S. W. Chee, J. M. Arce-Ramos, W. Li, A. Genest, U. Mirsaidov, Structural changes in noble metal nanoparticles during CO oxidation and their impact on catalyst activity. *Nat. Commun.* **11**, 2133 (2020). doi: 10.1038/s41467-020-16027-9.
5. S. Dai, Y. You, S. Zhang, W. Cai, M. Xu, L. Xie, R. Wu, G. W. Graham, X. Pan, In situ atomic-scale observation of oxygen-driven core-shell formation in $Pt_3Co$ nanoparticles. *Nat. Commun.* **8**, 204 (2017). doi: 10.1038/s41467-017-00161-y.
6. X. Zhang, S. Han, B. Zhu, G. Zhang, X. Li, Y. Gao, Z. Wu, B. Yang, Y. Liu, W. Baaziz, O. Ersen, M. Gu, J. T. Miller, W. Liu, Reversible loss of core–shell structure for Ni–Au bimetallic nanoparticles during $CO_2$ hydrogenation. *Nat. Catal.* **3**, 411–417 (2020). doi: 10.1038/s41929-020-0440-2.
7. A. R. Poerwoprajitno, L. Gloag, J. Watt, S. Cheong, X. Tan, H. Lei, H. A. Tahini, A. Henson, B. Subhash, N. M. Bedford, B. K. Miller, P. B. O'Mara, T. M. Benedetti, D. L. Huber, W. Zhang, S. C. Smith, J. J. Gooding, W. Schuhmann, R. D. Tilley, A single-Pt-atom-on-Ru-nanoparticle electrocatalyst for CO-resilient methanol oxidation. *Nat. Catal.* **5**, 231–237 (2022). doi: 10.1038/s41929-022-00756-9.
8. K. J.J. Mayrhofer, J. C. Meier, S. J. Ashton, G. K.H. Wiberg, F. Kraus, M. Hanzlik, M. Arenz, Fuel cell catalyst degradation on the nanoscale. *Electrochem. Commun.* **10**, 1144–1147 (2008). doi: 10.1016/j.elecom.2008.05.032.
9. R. M. Arán-Ais, Y. Yu, R. Hovden, J. Solla-Gullón, E. Herrero, J. M. Feliu, H. D. Abruña, Identical Location Transmission Electron Microscopy Imaging of Site-Selective Pt Nanocatalysts: Electrochemical Activation and Surface Disordering. *J. Am. Chem. Soc.* **137**, 14992–14998 (2015). doi: 10.1021/jacs.5b09553.
10. M. J. Williamson, R. M. Tromp, P. M. Vereecken, R. Hull, F. M. Ross, Dynamic microscopy of nanoscale cluster growth at the solid–liquid interface. *Nat. Mater.* **2**, 532–536 (2003). doi: 10.1038/nmat944.
11. Y. Yu, M. E. Holtz, H. L. Xin, D. Wang, H. D. Abruña, D. A. Muller, Understanding Pt-Co Catalyst Degradation Mechanism: from Ex-situ to In-situ. *Microsc. Microanal.* **19**, 1666–1667 (2013). doi: 10.1017/S1431927613010325.
12. L. Luo, M. H. Engelhard, Y. Shao, C. Wang, Revealing the Dynamics of Platinum Nanoparticle Catalysts on Carbon in Oxygen and Water Using Environmental TEM. *ACS Catal.* **7**, 7658–7664 (2017). doi: 10.1021/acscatal.7b02861.
13. Y. Yang, S. Louisia, S. Yu, J. Jin, I. Roh, C. Chen, M. V. Fonseca Guzman, J. Feijóo, P.-C. Chen, H. Wang, C. J. Pollock, X. Huang, Y.-T. Shao, C. Wang, D. A. Muller, H. D. Abruña, P. Yang, Operando studies reveal active Cu nanograins for $CO_2$ electroreduction. *Nature* **614**, 262–269 (2023). doi: 10.1038/s41586-022-05540-0.
14. Q. Zhang, Z. Song, X. Sun, Y. Liu, J. Wan, S. B. Betzler, Q. Zheng, J. Shangguan, K. C. Bustillo, P. Ercius, P. Narang, Y. Huang, H. Zheng, Atomic dynamics of electrified solid–

liquid interfaces in liquid-cell TEM. *Nature* **630**, 643–647 (2024). doi: 10.1038/s41586-024-07479-w.
15. D. Zhou, Y. Wang, H. H. Perez Garza, D. Su, An odyssey to operando environmental transmission electron microscopy: What's next? *Next. Mater.* **1**, 100007 (2023). doi: 10.1016/j.nxmate.2023.100007.
16. J. Miao, P. Ercius, S. J. L. Billinge, Atomic electron tomography: 3D structures without crystals. *Science* **353**, aaf2157 (2016). doi: 10.1126/science.aaf2157.
17. Y. Yang, J. Zhou, F. Zhu, Y. Yuan, D. J. Chang, D. S. Kim, M. Pham, A. Rana, X. Tian, Y. Yao, S. J. Osher, A. K. Schmid, L. Hu, P. Ercius, J. Miao, Determining the three-dimensional atomic structure of an amorphous solid. *Nature* **592**, 60–64 (2021). doi: 10.1038/s41586-021-03354-0.
18. Y. Zhang, Z. Li, X. Tong, Z. Xie, S. Huang, Y.-E. Zhang, H.-B. Ke, W.-H. Wang, J. Zhou, Three-dimensional atomic insights into the metal-oxide interface in Zr-$ZrO_2$ nanoparticles. *Nat. Commun.* **15**, 7624 (2024). doi: 10.1038/s41467-024-52026-w.
19. Z. Sun, Y. Zhang, Z. Li, Z. Xie, Y. Dai, X. Du, C. Ophus, J. Zhou, Strain release by 3D atomic misfit in fivefold twinned icosahedral nanoparticles with amorphization and dislocations. *Nat. Commun.* **16**, 1595 (2025). doi: 10.1038/s41467-025-56842-6.
20. S. Van Aert, K. J. Batenburg, M. D. Rossell, R. Erni, G. Van Tendeloo, Three-dimensional atomic imaging of crystalline nanoparticles. *Nature* **470**, 374–377 (2011). doi: 10.1038/nature09741.
21. B. Goris, S. Bals, W. Van den Broek, E. Carbó-Argibay, S. Gómez-Graña, L. M. Liz-Marzán, G. Van Tendeloo, Atomic-scale determination of surface facets in gold nanorods. *Nat. Mater.* **11**, 930–935 (2012). doi: 10.1038/nmat3462.
22. L. Jones, K. E. MacArthur, V. T. Fauske, A. T. J. van Helvoort, P. D. Nellist, Rapid Estimation of Catalyst Nanoparticle Morphology and Atomic-Coordination by High-Resolution Z-Contrast Electron Microscopy. *Nano Lett.* **14**, 6336–6341 (2014). doi: 10.1021/nl502762m.
23. A. De Backer, K. H. W. van den Bos, W. Van den Broek, J. Sijbers, S. Van Aert, StatSTEM: An efficient approach for accurate and precise model-based quantification of atomic resolution electron microscopy images. *Ultramicroscopy* **171**, 104–116 (2016). doi: 10.1016/j.ultramic.2016.08.018.
24. J. Park, H. Elmlund, P. Ercius, J. M. Yuk, D. T. Limmer, Q. Chen, K. Kim, S. H. Han, D. A. Weitz, A. Zettl, A. P. Alivisatos, Nanoparticle imaging. 3D structure of individual nanocrystals in solution by electron microscopy. *Science* **349**, 290–295 (2015). doi: 10.1126/science.aab1343.
25. B. H. Kim, J. Heo, S. Kim, C. F. Reboul, H. Chun, D. Kang, H. Bae, H. Hyun, J. Lim, H. Lee, B. Han, T. Hyeon, A. P. Alivisatos, P. Ercius, H. Elmlund, J. Park, Critical differences in 3D atomic structure of individual ligand-protected nanocrystals in solution. *Science* **368**, 60–67 (2020). doi: 10.1126/science.aax3233.
26. S. Kang, J. Kim, S. Kim, H. Chun, J. Heo, C. F. Reboul, R. Meana-Pañeda, C. T. S. Van, H. Choi, Y. Lee, J. Rhee, M. Lee, D. Kang, B. H. Kim, T. Hyeon, B. Han, P. Ercius, W. C. Lee, H. Elmlund, J. Park, Time-resolved Brownian tomography of single nanocrystals in liquid during oxidative etching. *Nat. Commun.* **16**, 1158 (2025). doi: 10.1038/s41467-025-56476-8.
27. Y. Yang, J. Zhou, Z. Zhao, G. Sun, S. Moniri, C. Ophus, Y. Yang, Z. Wei, Y. Yuan, C. Zhu, Y. Liu, Q. Sun, Q. Jia, H. Heinz, J. Ciston, P. Ercius, P. Sautet, Y. Huang, J. Miao, Atomic-scale identification of active sites of oxygen reduction nanocatalysts. *Nat. Catal.* **7**, 796–806 (2024). doi: 10.1038/s41929-024-01175-8.

28. Y. Dai, Z. Xie, Y. Zhang, X. Du, Z. Li, J. Xie, Z. Sun, J. Zhou, Mapping Surface and Subsurface Atomic Structures of Au@Pd Core-Shell Nanoparticles in Three Dimensions. *ACS Nano* **19**, 9006–9016 (2025). doi: 10.1021/acsnano.4c17462.
29. C. Jeong, J. Lee, H. Jo, K. Lee, S. Lee, C. Ophus, P. Ercius, E. Cho, Y. Yang, Atomic-scale 3D structural dynamics and functional degradation of Pt alloy nanocatalysts during the oxygen reduction reaction. *Nat. Commun.* **16**, 8026 (2025). doi: 10.1038/s41467-025-63448-5.
30. A. De Backer, L. Jones, I. Lobato, T. Altantzis, B. Goris, P. D. Nellist, S. Bals, S. Van Aert, Three-dimensional atomic models from a single projection using Z-contrast imaging: verification by electron tomography and opportunities. *Nanoscale* **9**, 8791–8798 (2017). doi: 10.1039/c7nr02656k.
31. D. Kang, S. Kim, J. Heo, D. Kim, H. Bae, S. Kang, S. Shim, H. Lee, J. Park, Complex ligand adsorption on 3D atomic surfaces of synthesized nanoparticles investigated by machine-learning accelerated ab initio calculation. *Nanoscale* **15**, 532–539 (2023). doi: 10.1039/D2NR05294F.
32. Y. Yang, C.-C. Chen, M. C. Scott, C. Ophus, R. Xu, A. Pryor, L. Wu, F. Sun, W. Theis, J. Zhou, M. Eisenbach, P. R. C. Kent, R. F. Sabirianov, H. Zeng, P. Ercius, J. Miao, Deciphering chemical order/disorder and material properties at the single-atom level. *Nature* **542**, 75–79 (2017). doi: 10.1038/nature21042.
33. X. Tian, D. S. Kim, S. Yang, C. J. Ciccarino, Y. Gong, Y. Yang, Y. Yang, B. Duschatko, Y. Yuan, P. M. Ajayan, J. C. Idrobo, P. Narang, J. Miao, Correlating the three-dimensional atomic defects and electronic properties of two-dimensional transition metal dichalcogenides. *Nat. Mater.* **19**, 867–873 (2020). doi: 10.1038/s41563-020-0636-5.
34. S. Moniri, Y. Yang, J. Ding, Y. Yuan, J. Zhou, L. Yang, F. Zhu, Y. Liao, Y. Yao, L. Hu, P. Ercius, J. Miao, Three-dimensional atomic structure and local chemical order of medium- and high-entropy nanoalloys. *Nature* **624**, 564–569 (2023). doi: 10.1038/s41586-023-06785-z.
35. J. Zhou, Y. Yang, Y. Yang, D. S. Kim, A. Yuan, X. Tian, C. Ophus, F. Sun, A. K. Schmid, M. Nathanson, H. Heinz, Q. An, H. Zeng, P. Ercius, J. Miao, Observing crystal nucleation in four dimensions using atomic electron tomography. *Nature* **570**, 500–503 (2019). doi: 10.1038/s41586-019-1317-x.
36. B. Lim, M. Jiang, P. H. C. Camargo, E. C. Cho, J. Tao, X. Lu, Y. Zhu, Y. Xia, Pd-Pt Bimetallic Nanodendrites with High Activity for Oxygen Reduction. *Science* **324**, 1302–1305 (2009). doi: 10.1126/science.1170377.
37. L. Zhang, L. T. Roling, X. Wang, M. Vara, M. Chi, J. Liu, S.-I. Choi, J. Park, J. A. Herron, Z. Xie, M. Mavrikakis, Y. Xia, Platinum-based nanocages with subnanometer-thick walls and well-defined, controllable facets. *Science* **349**, 412–416 (2015). doi: 10.1126/science.aab0801.
38. T. He, W. Wang, F. Shi, X. Yang, X. Li, J. Wu, Y. Yin, M. Jin, Mastering the surface strain of platinum catalysts for efficient electrocatalysis. *Nature* **598**, 76–81 (2021). doi: 10.1038/s41586-021-03870-z.
39. L. Tao, K. Wang, F. Lv, H. Mi, F. Lin, H. Luo, H. Guo, Q. Zhang, L. Gu, M. Luo, S. Guo, Precise synthetic control of exclusive ligand effect boosts oxygen reduction catalysis. *Nat. Commun.* **14**, 6893 (2023). doi: 10.1038/s41467-023-42514-w.
40. M. Li, K. Duanmu, C. Wan, T. Cheng, L. Zhang, S. Dai, W. Chen, Z. Zhao, P. Li, H. Fei, Y. Zhu, R. Yu, J. Luo, K. Zang, Z. Lin, M. Ding, J. Huang, H. Sun, J. Guo, X. Pan, W. A. Goddard, P. Sautet, Y. Huang, X. Duan, Single-atom tailoring of platinum nanocatalysts for high-performance multifunctional electrocatalysis. *Nat. Catal.* **2**, 495–503 (2019). doi: 10.1038/s41929-019-0279-6.
41. J. Chang, G. Wang, M. Wang, Q. Wang, B. Li, H. Zhou, Y. Zhu, W. Zhang, M. Omer, N. Orlovskaya, Q. Ma, M. Gu, Z. Feng, G. Wang, Y. Yang, Improving Pd–N–C fuel cell electrocatalysts through fluorination-driven rearrangements of local coordination environment. *Nat. Energy* **6**, 1144–1153 (2021). doi: 10.1038/s41560-021-00940-4.

42. J. Chang, G. Wang, C. Li, Y. He, Y. Zhu, W. Zhang, M. Sajid, A. Kara, M. Gu, Y. Yang, Rational design of septenary high-entropy alloy for direct ethanol fuel cells. *Joule* **7**, 587–602 (2023). doi: 10.1016/j.joule.2023.02.011.
43. Z. Li, Z. Xie, Y. Zhang, X. Mu, J. Xie, H.-J. Yin, Y.-W. Zhang, C. Ophus, J. Zhou, Probing the atomically diffuse interfaces in Pd@Pt core-shell nanoparticles in three dimensions. *Nat. Commun.* **14**, 2934 (2023). doi: 10.1038/s41467-023-38536-z.
44. J. Xie, Z. Xie, Z. Li, Y. Dai, Y. Zhang, Y. Bu, C. Liu, H. Jiang, M. Li, J. Zhou, Revealing the interface-driven atomic local chemical heterogeneity in bimetallic catalysts in three dimensions. Research Square [Preprint] (2024). https://doi.org/10.21203/rs.3.rs-4185973/v1.
45. D. A. Porter, K. E. Easterling, M. Y. Sherif, Phase Transformations in Metals and Alloys (CRC, 2021).
46. H. Shan, W. Gao, Y. Xiong, F. Shi, Y. Yan, Y. Ma, W. Shang, P. Tao, C. Song, T. Deng, H. Zhang, D. Yang, X. Pan, J. Wu, Nanoscale kinetics of asymmetrical corrosion in core-shell nanoparticles. *Nat. Commun.* **9**, 1011 (2018). doi: 10.1038/s41467-018-03372-z.
47. F. Shi, W. Gao, H. Shan, F. Li, Y. Xiong, J. Peng, Q. Xiang, W. Chen, P. Tao, C. Song, W. Shang, T. Deng, H. Zhu, H. Zhang, D. Yang, X. Pan, J. Wu, Strain-Induced Corrosion Kinetics at Nanoscale Are Revealed in Liquid: Enabling Control of Corrosion Dynamics of Electrocatalysis. *Chem* **6**, 2257–2271 (2020). doi: 10.1016/j.chempr.2020.06.004.
48. F. Shi, P. Tieu, H. Hu, J. Peng, W. Zhang, F. Li, P. Tao, C. Song, W. Shang, T. Deng, W. Gao, X. Pan, J. Wu, Direct in-situ imaging of electrochemical corrosion of Pd-Pt core-shell electrocatalysts. *Nat. Commun.* **15**, 5084 (2024). doi: 10.1038/s41467-024-49434-3.
49. H. L. Skriver, N. M. Rosengaard, Surface energy and work function of elemental metals. *Phys. Rev. B* **46**, 7157–7168 (1992). doi: 10.1103/physrevb.46.7157.
50. S. Xue, G. Chen, F. Li, Y. Zhao, Q. Zeng, J. Peng, F. Shi, W. Zhang, Y. Wang, J. Wu, R. Che, Understanding of Strain-Induced Electronic Structure Changes in Metal-Based Electrocatalysts: Using Pd@Pt Core-Shell Nanocrystals as an Ideal Platform. *Small* **17**, 2100559 (2021). doi: 10.1002/smll.202100559.
51. K. Dabov, A. Foi, V. Katkovnik, K. Egiazarian, Image Denoising by Sparse 3-D Transform-Domain Collaborative Filtering. *IEEE Trans. on Image Process.* **16**, 2080–2095 (2007). doi: 10.1109/TIP.2007.901238.
52. N. Otsu, A Threshold Selection Method from Gray-Level Histograms. *IEEE Trans. Syst. Man Cybern.* **9**, 62–66 (1979). doi: 10.1109/TSMC.1979.4310076.
53. M. Pham, Y. Yuan, A. Rana, S. Osher, J. Miao, Accurate real space iterative reconstruction (RESIRE) algorithm for tomography. *Sci. Rep.* **13**, 5624 (2023). doi: 10.1038/s41598-023-31124-7.
54. S. S. Rogers, T. A. Waigh, X. Zhao, J. R. Lu, Precise particle tracking against a complicated background: polynomial fitting with Gaussian weight. *Phys. Biol.* **4**, 220–227 (2007). doi: 10.1088/1478-3975/4/3/008.
55. M. C. Scott, C.-C. Chen, M. Mecklenburg, C. Zhu, R. Xu, P. Ercius, U. Dahmen, B. C. Regan, J. Miao, Electron tomography at 2.4-ångström resolution. *Nature* **483**, 444–447 (2012). doi: 10.1038/nature10934.
56. C.-C. Chen, C. Zhu, E. R. White, C.-Y. Chiu, M. C. Scott, B. C. Regan, L. D. Marks, Y. Huang, J. Miao, Three-dimensional imaging of dislocations in a nanoparticle at atomic resolution. *Nature* **496**, 74–77 (2013). doi: 10.1038/nature12009.
57. A. T. Brünger, P. D. Adams, G. M. Clore, W. L. DeLano, P. Gros, R. W. Grosse-Kunstleve, J.-S. Jiang, J. Kuszewski, M. Nilges, N. S. Pannu, R. J. Read, L. M. Rice, T. Simonson, G. L.

Warren, Crystallography & NMR System: A New Software Suite for Macromolecular Structure Determination. *Acta Crystallogr., Sect. D: Biol. Crystallogr.* **D54**, 905–921 (1998). doi: 10.1107/s0907444998003254.
58. S. Lloyd, Least squares quantization in PCM. *IEEE Trans. Inf. Theory* **28**, 129–137 (1982). doi: 10.1109/TIT.1982.1056489.
59. A. Stukowski, Computational Analysis Methods in Atomistic Modeling of Crystals. *JOM* **66**, 399–407 (2014). doi: 10.1007/s11837-013-0827-5.
60. C. L. Kelchner, S. J. Plimpton, J. C. Hamilton, Dislocation nucleation and defect structure during surface indentation. *Phys. Rev. B* **58**, 11085–11088 (1998). doi: 10.1103/PhysRevB.58.11085.
61. R. Xu, C.-C. Chen, L. Wu, M. C. Scott, W. Theis, C. Ophus, M. Bartels, Y. Yang, H. Ramezani-Dakhel, M. R. Sawaya, H. Heinz, L. D. Marks, P. Ercius, J. Miao, Three-dimensional coordinates of individual atoms in materials revealed by electron tomography. *Nat. Mater.* **14**, 1099–1103 (2015). doi: 10.1038/nmat4426.
62. Q.-J. Li, H. Sheng, E. Ma, Strengthening in multi-principal element alloys with local-chemical-order roughened dislocation pathways. *Nat. Commun.* **10**, 3563 (2019). doi: 10.1038/s41467-019-11464-7.

**Acknowledgments:** The authors thank the Analysis Center at Tsinghua University for the use of aberration-corrected electron microscopes, The authors thank the Analytical Instrumentation Center at Peking University for the use of ICP-AES. The authors thank CHIPNOVA company for the use of in situ electrochemical TEM holder. This work is also supported by High-performance Computing Platform of Peking University. The authors thank S. Zhou for the discussion in the pre-exploration for demonstration of 4D-ecAET in 2D.

**Funding:** This work was supported by the National Key R&D Program of China (Grant No. 2024YFA1509500), the National Natural Science Foundation of China (Grant No. 22172003, 92477203 & 22372004) and Beijing Natural Science Foundation (Grant No. F251007).

**Author contributions:** J.Z. and J.X. conceived the idea, J.Z and M. L. directed the study. J.X. conducted the four-dimensional atomic-resolution ET data collection experiments under the supervision of J.Z. J.X. and D.J. synthesized the Pd@Pt core-shell nanoparticles. J.X. implemented the electrochemical measurements under the supervision of M.L., S.L. and Y.D. J.X. and Z.X. carried out the experiments for demonstration of 4D-ecAET in 2D. J.X., Z.X., Y.D. and Y.Z. discussed the methods of atomic-resolution ET reconstruction. J.X. analyzed the 3D atomic structures and interpreted the results under the supervision of J.Z. J.X. and S. L. completed in situ electrochemical TEM tests under the assistance of CHIPNOVA company. J.X. and J.Z. wrote the manuscript. All authors commented on the manuscript.

**Competing interests:** Authors declare that they have no competing interests.

**Data and materials availability:** All data are available in the main text or the supplementary materials.

**Supplementary Materials**

Materials and Methods

Figs. S1 to S29; Tables S1 and S3

Movies S1 to S5

**Materials and Methods**

1. Chemicals

Sodium tetrachloropalladate(II) ($Na_2PdCl_4$, ≥99.99%) was purchased from Sigma-Aldrich; chloroplatinic acid hexahydrate ($H_2PtCl_6·6H_2O$, 99.7%), D-glucose, ethanol, cyclohexane and acetone were purchased from tgchem, Beijing, China; potassium hydroxide (KOH, 99.7%) was purchased from Greagent; potassium bromide (KBr, 99.7%) and Nafion 117 solution were purchased from MREDA, Beijing, China; potassium chloride (KCl, 99.7%) and L-ascorbic acid (99.7%) were purchased from Xilong Scientifi, Guangdong, China; Poly(vinyl-pyrrolidone) (PVP, MW ≈ 55,000 g/mol) was purchased from Yuanye, Shanghai, China; oleylamine (90%, No. 1017269) was purchased from Leyan, Shanghai, China; hydroxy purified multi-walled carbon nanotube (CNT, outer diameter: 8-15 nm, length: 0.5-2 μm) was purchased from Aladdin, Shanghai, China. All the chemicals were used as received without further purification.

2. Sample preparation

Pd@Pt nanoparticles (NPs) were synthesized by a modified method described in detail elsewhere (*44, 50*).

1) *Synthesis of Pd NPs*

In a typical synthesis, 315 mg PVP, 15 mg KBr, 610.5 mg KCl and 180 mg L-ascorbic acid were dissolved in 24 mL of deionized water and heated to 80 °C. 30 mg $Na_2PdCl_4$ was dissolved in 9 mL deionized water, ultrasonicated for 10 min, and mixed with the pre-heated solutions. The mixture was heated at 80 °C for 4 h. The products were collected after adding 5 mL ethanol and 33 mL acetone and following by a 2-min centrifugation under 10000 rpm. The products were re-dispersed in 6 mL ethanol-water mixture (1:1, v:v), and ultrasonicated for 5min for further use.

2) *Synthesis of Pd@Pt NPs*

6 mL Pd NPs from above were mixed with 6 mL oleylamine, stirred at 200 °C for 5 min to remove ethanol. After cooling to room temperature, the solutions were mixed with 75 mg glucose, 6 mL of oleylamine, and 0.750 mL of 10 mg mL$^{-1}$ $H_2PtCl_6$ • $6H_2O$-oleylamine solution in a glass vial. The mixture was heated at 200 ℃ for 30 min. The products were washed 3 times with 3 mL cyclohexane and 35 mL ethanol, re-dispersed in cyclohexane for further use.

3) *Pd@Pt/CNT catalyst loading and ink preparation*

Pd@Pt NPs were mixed with CNT, dispersed in ethanol and ultrasonicated for 30 min. The mixture was stirred for 12 h and centrifuged to get raw products. The precipitation was cleaned by stirring in the solutions of ethanol (10 mL) and acetic acid (10 mL) for another 1 h. The product was further washed with ethanol, and dried for further use. The actual Pt and Pd loadings of catalysts were determined by inductively coupled plasma - atomic emission spectrometry (ICP-AES) to be 3.1 wt% and 10.1 wt%, repectively. The catalyst inks were prepared by mixing catalyst powders with ethanol and 5 wt% Nafion 117 solution (the volume ratio is 49:1) and ultrasonicating for 30 min. The concentration of catalysts in inks is controlled to be 1.44 mg mL$^{-1}$.

3. General material characterizations

Annular dark-field scanning transmission electron microscopy (ADF-STEM) and energy-dispersive X-ray spectroscopy (EDS) tests were carried out on an aberration-corrected FEI

Themis Z microscope, operating at 300 kV. For the test of EDS mapping of catalysts after 2,000 cycles (fig. S22), 20 μL inks were drop-cast onto a carbon paper (the size is 1cm*2 cm), and dried naturally as the working electrode. The quantitative elemental compositions of precious metals in all catalysts were determined by ICP-AES (Leeman Prodigy 7 ICP-AES).

3. Experimental details in four-dimensional electrocatalytic atomic-resolution tomography
   1) *General work flow*
   This section only introduces the overall workflow so that readers can quickly understand the experiment. The details of experiments mentioned in this section will be explained in subsequent sections.
   Firstly, one tomographic tilt series of Pd@Pt was acquired on a prepared specimen grid. Lower-magnification images with position information of the targeted NP (e.g., the shape of the carbon nanotubes surrounded and the relative position of NPs surrounded) were recorded (fig. S3). Based on the information, same targeted NPs could be found around the identical positions after the electrochemical test, as long as the part of membrane is still intact.
   Next, the grid was clamped by a glassy carbon electrode clamp to serve as a working electrode. A durability test was carried out under the working condition of electrocatalysis. After the reaction, the grid was washed by water-ethanol mixture to remove the residual alkali, dried in vacuum and kept in an Argon-filled glovebox for the next acquisition of tilt dataset.
   The process was repeated to get the whole set of data series.

   2) *Specimen grid / work electrode preparation*
   A 200-mesh formvar carbon gold grid purchased from Electron Microscopy Sciences was used, for the 200-mesh gold skeleton owns good conductivity, better supportiveness and no reaction activity at the working condition in this study (fig. S2). To further enhance the conductivity of the membrane, a 4-nm thick carbon layer was pre-deposited onto the grid through a sputter coater (pulse mode in Leica EM ACE600). Then the grid was heated at 350 °C for 12 h in an Argon-filled glove box to remove contamination.
   A certain amount of ink (table S1) containing Pd@Pt/CNT was drop-cast onto the edges the grid from above and again sprayed onto the grid using an atomizer. Then the grid was dried in the air, kept in an Argon-filled glove box for the further use.

   3) *Acquisitions of tomographic tilt series*
   Tomographic tilt series on the same NPs were acquired in ADF-STEM mode on an aberration-corrected FEI Themis Z microscope, the operation parameters of datasets are 300 kV voltage, 25 mrad convergence semi-angle, 40.6 mrad HAADF detector inner semi-angle and 200 mrad HAADF detector outer semi-angle, and the pixel size is 0.352 Å. To minimize the sample drift, three sequential frames of images per projections were acquired with 2 μs dwell time in each tilt angle.
   A much lower electron dose rate was used to protect the NPs from beam damage. For each tomographic tilt series, the electron dose rate was estimated to be between $2.5 \times 10^5$ e$^-$ Å$^{-2}$ and $3.0 \times 10^5$ e$^-$ Å$^{-2}$ (table S2), which is lower than 40% of that used in ref. (*35*).
   To check if there are any potential structural changes induced by the electron beam irradiation, we compared all the experimental zero-degree images with forward projection images of final three-dimensional (3D) atomic models (*43*). No noticeable structural changes were observed during the data acquisition for particles 1, 2 and 3.

   4) Electrochemical measurements

The prepared grid was used as the working electrode. A Hg/HgO electrode was used as the reference electrode, a platinum wire electrode was used as the counter electrode. The durability test was carried out by cyclic voltammetry (CV), which was performed at 0.05-1.1 V versus RHE ($V_{RHE}$) in Argon-saturated 1 M ethanol/KOH solution. The scan rate was set to be 50 mV s$^{-1}$. Different cycles (i.e., 100, 200, 500, 1,500 and 2,500 cycles) were performed to get nanocatalysts at different periods throughout the life cycles.

4. 3D atomic-resolution reconstruction
   1) *Image pre-processing*
   For each acquired tilt series, the images were pre-processed by using the method described elsewhere (*34, 43*) before reconstruction: First, the frames of images at each angle were registered by normalized cross-correlation and then averaged to enhance signal-to-noise ratio. Linear drifts are estimated and corrected during image registration. Next, the block-matching and 3D filtering algorithm (*51*) was used to remove the Poisson and Gaussian noise from all registered images. Then, for each denoised image, a mask of NP was generated based on Otsu thresholding (*52*). To avoid losing any surface information, generally a certain distance about 10 pixels was left between the mask and the particles. We used a Laplace interpolation to estimate the background value within the mask and then subtracted it from the image. Finally, the center-of-mass alignment and common-line alignment were used to align the images in each tilt dataset with a sub-pixel accuracy.

   2) *3D reconstruction*
   Real Space Iterative Reconstruction (RESIRE) algorithm (*53*) was used to reconstruct each experimental tilt series (Parameters used are recorded in table S2). Spatial re-alignments and angular refinements were performed to reduce the angle bias caused by the instability of the stage and holder. The background of image was re-evaluated and re-subtracted, following by further spatial re-alignments and angular refinements. The resulting projections and angles were used to generate the final 3D volume by RESIRE.

5. Determination of 3D atomic coordinates and chemical species
   Using the final 3D volume from above, a series of steps were carried out to get the atomic coordinates and chemical.

   1) *Identification of local intensity maxima*
   Polynomial fitting method (*17, 54*) were performed within a $7 \times 7 \times 7$ voxel box to locate the peak position. Based on the initial list, each local maximum would be recognized as a potential atomic coordinate with the constraint that the minimum distance between the neighboring peaks cannot be lower than 2.2 Å. 3D polynomial fitting was performed on each potential atom coordinate to get sub-pixel resolution. In this step, more than 98% of the final atomic coordinates are automatically searched and determined. Due to the "missing wedge" problem (*55, 56*), artifacts and noises exist in the final reconstructions, resulting fitting errors and false atomic sites. The results from above were further checked for possible unidentified and misidentified atoms due to the error. A small fraction of atoms are adjusted within manually-given regions (based on the local intensity maxima), which is routinely used during the atom tracing and refinement process in macro-molecular crystallography (*57*).

   2) *Element classification*

Element classification is performed by means of the K-means clustering algorithm (*58*) to get a raw result of classification. The integrated intensity sums in the range of a $7 \times 7 \times 7$ voxel box near each atomic coordinate. All integrated intensities were classified into two categories, Pd and Pt atoms. Owing to the "missing wedge" problem (*55, 56*), the intensities located at the surface of the NPs are weaker than the intensities located inside the particles in the direction of the "missing wedge". As a result, some surface Pt atoms are misclassified into Pd. We performed a previously reported method called "local re-classification" (*27, 35*) to alleviate the effect of the intensity variation: Integrated intensities of atoms are re-classified by K-means algorithm. For each atom as a center position, we cropped a sphere with a radius of 6.75 Å, the average intensity distribution of the two types was computed within the sphere. The integrated intensity of the center atom was compared with that of the average Pd and Pt atom, then the center atom was assigned to the type with a closer distance of integrated intensity. The process was iterated for all atoms until the reclassification procedure converged. For Pd@Pt$_{3\_2.5K}$ with a lower Pt content, we identified a small fraction of atoms with strongest integrated intensities as Pt atoms, and then use the local re-classification method to recover the misclassified atoms.

Final atomic models containing 3D atomic coordinates and chemical species are used for further structural analysis, the detailed structural information is recorded in Table S2.

6. Structural analysis of final atomic models
1) *Calculation of CN, specific surface area, distance to surface and surface facet distribution*
Pair distribution functions (PDF) were calculated based on the atomic coordinates. Using first valleys in PDFs as the threshold, CN was calculated. Atoms with CN < 12 was classified as surface atoms. The specific surface area were calculated by measuring the area of a surface mesh generated by the alpha-shape algorithm (*59*). First peaks in PDFs were used as the probe sphere radius parameter. Similarly, distances to surface were calculated by measuring the distances of each atom to the closest point on the surface mesh. For the calculation of surface facet distribution, the fractions of {111}, {110} and {100} atoms were classified based on information of both CN and centrosymmetry parameter (CSP) (*60*). CSP is used to measure the degree of the local lattice disorder around an atom. The contribution of all neighbor pairs around an atom in a perfect lattice would lead to a lower value CSP. Edges, corners and other high-index facets own higher values of CSP, which are defined as the "other" category in this study.

2) *Comparisons of common positions and consistent atoms*
First, an ideal fcc lattice was estimated by a least-squares fit of the atom positions for each datasets (*61*), all of the atoms are fitted successfully. Next, we compared two atomic models of the same NP with each other. We identified pairs of atoms (that is, one atom from each dataset to form a pair) with deviations smaller than half of the first valley in PDFs (*35*). The pairs of atoms between two datasets are recognized as common positions (Table S2). The positions that did not form a pair during this process are recognized as disappearing or appearing atoms in the dataset. Then, we calculated the consistency of atoms. Most of the atom pairs have the same atomic species, which are recognized as the consistent atoms / the core (Table S2).
3) Analysis of Pd/Pt heterogeneity
Local chemical heterogeneity was calculated within the 4$^{th}$ shell (corresponding to two layers of neighboring atoms), which was recently found to be functional in modulating electronic structures of catalysts (*44*). One central atom that owns a coordination shell consisted of both Pd and Pt, is recognized as a heterogeneous center. The Pd concentration ($c_{Pd}$) within the coordination shell was then quantified for heterogeneous Pt centers.

4) Chemical short-range-order parameter (CSROP)

The short-range-order parameter of each NP was calculated via the pair-wise multicomponent short-range-order parameter (62), which is defined as $\alpha_{ij}^m = \left(p_{ij}^m - C_j\right) / \left(\delta_{ij} - C_j\right)$, where $m$ means the $m$th nearst-neighbor shell of the central atom $i$, $p_{ij}^m$ is the average probability of finding a $j$-type atom around an $i$-type atom in the $m$th shell, $C_j$ is the average concentration of $j$-type atom in the system, and $\delta_{ij}$ is the Kronecker delta function. In this study, $\alpha_{Pd-Pd}$ and $\alpha_{Pt-Pd}$ are calculated ($m = 4$). For pairs of the same species (i.e., $\alpha_{Pd-Pd}$), a positive $\alpha_{Pd-Pd}$ suggests the tendency of Pd segregation in the shell, and a negative $\alpha_{Pd-Pd}$ means the opposite. In contrast, for pairs of different elements (i.e., $\alpha_{Pt-Pd}$), a negative $\alpha_{Pt-Pd}$ means the tendency of Pd clustering in the shell of central Pt atom, while a positive $\alpha_{Pt-Pd}$ means the opposite. $-\alpha_{Pt-Pd}$ and $\alpha_{Pd-Pd}$ were used to investigate the evolution of degrees of Pd segregation towards different periods of rection, for both of them show Pd-abundant tendency when the values are positive.

The surface atoms were classified into 8 regions of {111} orientations (upper panel) and 6 regions of {100} orientations (lower panel), based on the information of facet distribution from fig. S16. Then the averages of CSROPs were calculated.

7. In situ electrochemical TEM

In situ electrochemical liquid phase TEM experiments were carried out on the Chip-Nova in situ EC-TEM holder in an FEI Talos F200S microscope with an acceleration voltage of 200 kV. Two silicon wafers were used as chips, in which a 10-nm-thick $SiN_x$ window, Au working electrode and counter/reference electrodes were constructed in the bottom chip. The distance between Au working electrode and counter/reference electrodes is 200 μm. The Au reference electrode was calibrated by measuring the same CV curves of Pd@Pt catalysts using the reversible hydrogen electrode (RHE) and Au plate as reference electrodes as previous report (48). In the same electrolyte, $E_{RHE} = E_{Au} + 0.88$ V. Pd@Pt NPs were dropped onto the Au electrodes of the bottom chip, then the bottom and top chips were assembled for further use. The solution containing 0.1 M KOH and 0.1 M ethanol were used as electrolyte. An electric field was applied on the liquid cell through the CHI 660E electrochemical workstation.

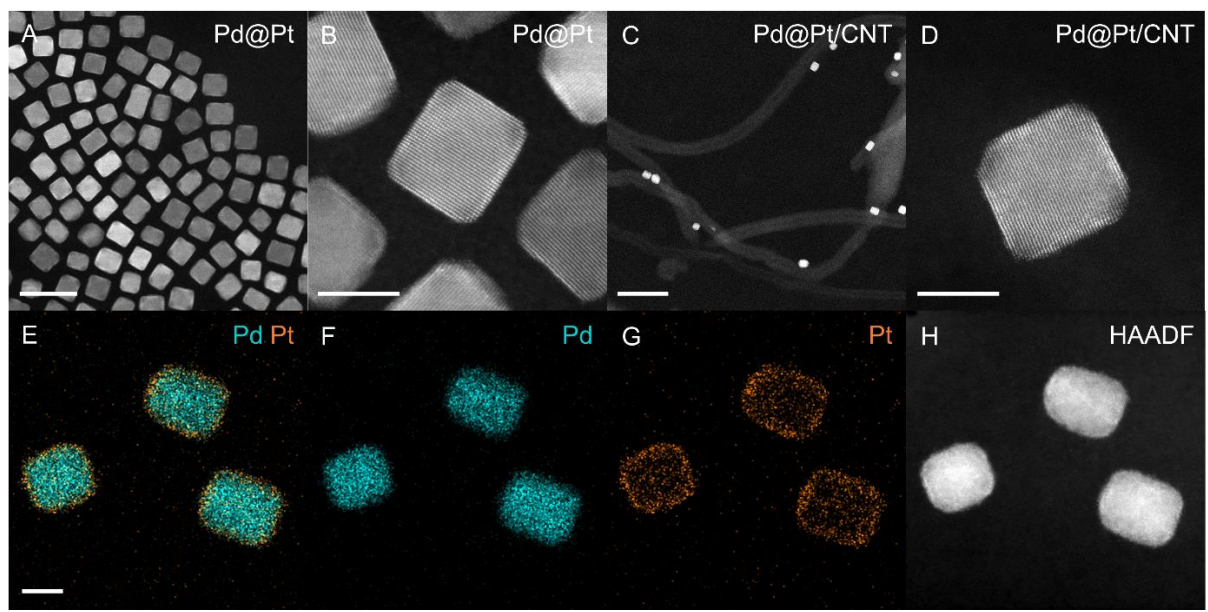

**Fig. S1. Characterization of Pd@Pt NPs and Pd@Pt/CNT nanocatalysts.**
(**A** and **B**) Atomic resolution ADF-STEM images of Pd@Pt NPs at (A) low and (B) high magnifications. Scale bar is 20 nm for (A) and 5 nm for (B). (**C** and **D**) Atomic resolution ADF-STEM images of Pd@Pt/CNT nanocatalysts at (C) low and (D) high magnifications. Scale bar is 50 nm for (C) and 5 nm for (D). (**E** to **H**) EDS mapping of Pd@Pt/CNT NPs: images of (E) merged Pd and Pt signals, (F) Pd, (G) Pt and (H) ADF signals. Scale bars are 5nm for (E) to (H). Both Pd@Pt NPs and Pd@Pt/CNT show a cuboctahedral shape and are in good uniformity.

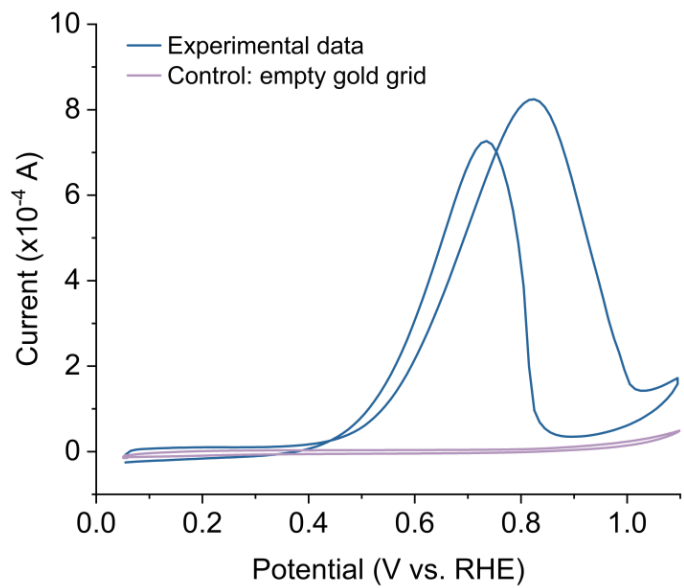

**Fig. S2. CV curves tested with a prepared sample and an empty gold grid.**

While the experimental data shows a typical electrochemical curve of electrocatalytic ethanol oxidation reaction, the curve tested on an empty gold grid without any catalyst shows no reaction activity towards the working condition in this study.

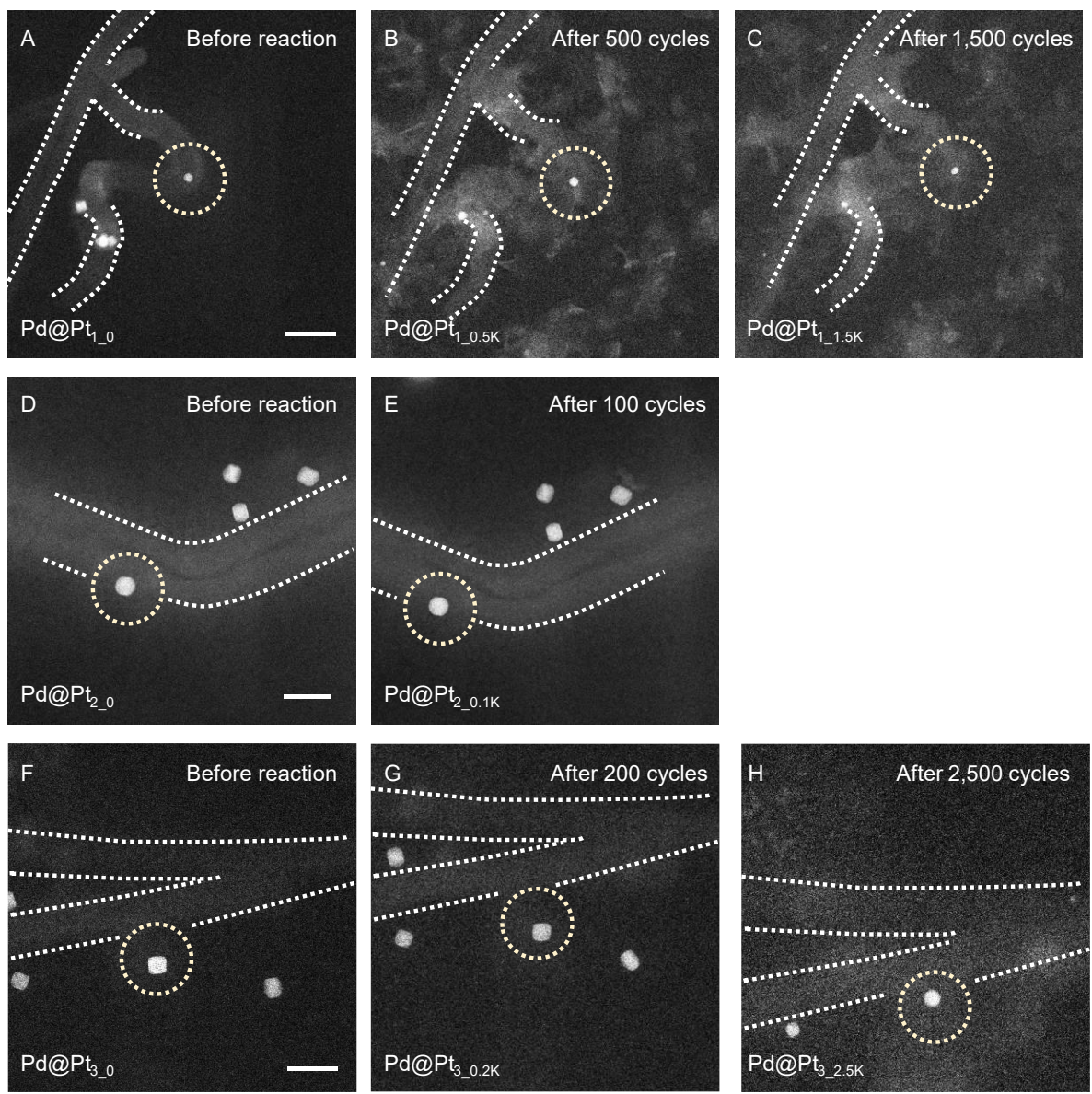

**Fig. S3. Images around the same NPs before and after reactions.**
(**A** to **C**) Low-magnification ADF-STEM images of particle-1 recorded (A) before reaction, (B) after 500 cycles and (C) after 1,500 cycles. Scale bars is 50 nm for (A) to (C). (**D** and **E**) Low-magnification ADF-STEM images of particle-2 recorded (D) before reaction and (E) after 100 cycles. (**F** to **H**) Low-magnification ADF-STEM images of particle-3 recorded (F) before reaction, (G) after 200 cycles and (H) after 2500 cycles. Targeted NPs are in the center of yellow dashed circles. The intensities lighter than NPs show the shapes of surrounding carbon nanotubes (white dashed curves), which kept intact after different cycles. Scale bars are 20 nm for (D) to (H).

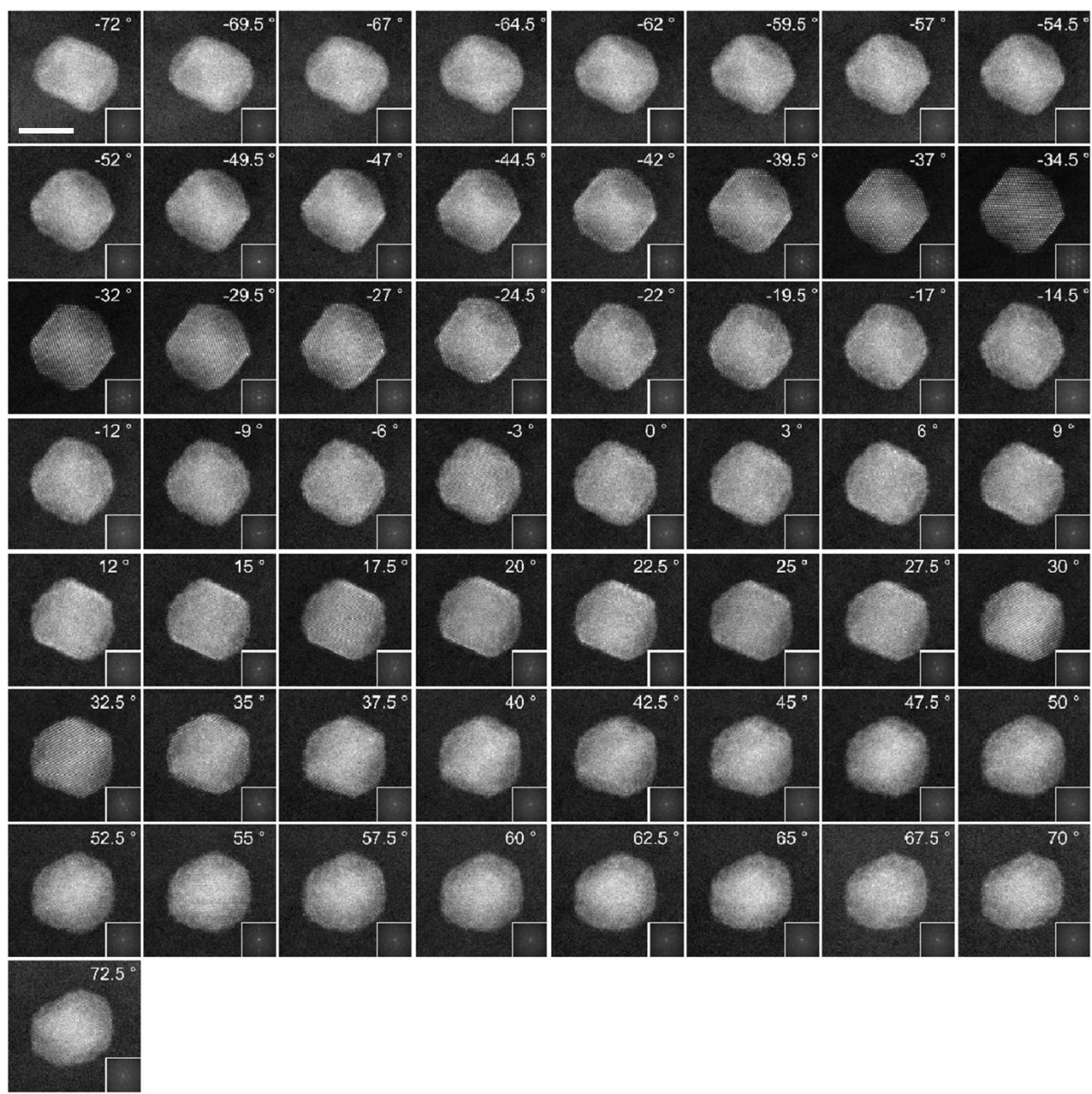

**Fig. S4. Tomographic tilt series of Pd@Pt$_{1\_0}$ NP.**
ADF-STEM images with a tilting range from −72.0° to +72.5°. The value of angle to which the image belongs is recorded in the upper right corner of each image. Scale bar is 5 nm.

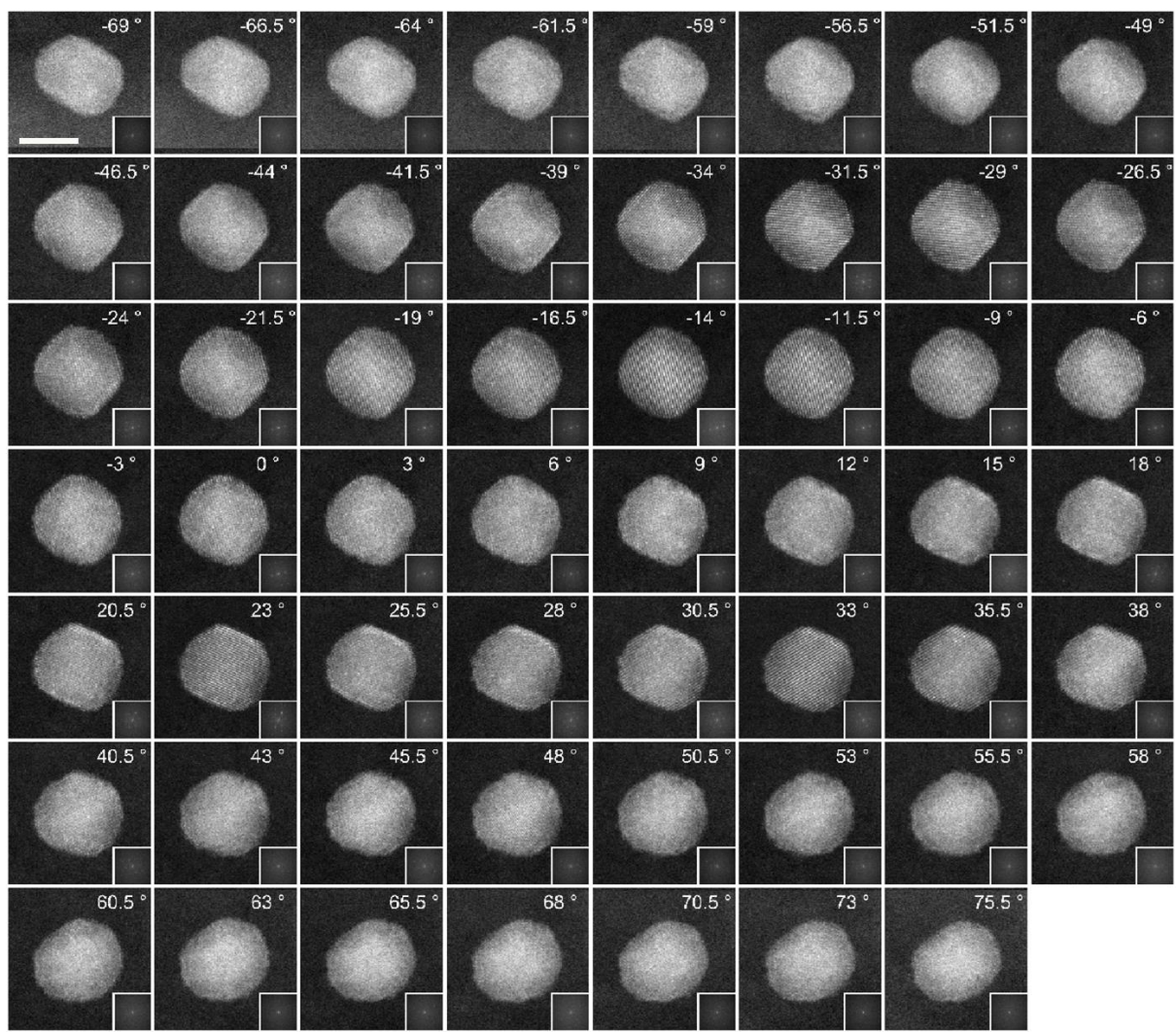

**Fig. S5. Tomographic tilt series of Pd@Pt$_{1\_0.5K}$ NP.**
ADF-STEM images with a tilting range from −69.0° to +75.5°. The value of angle to which the image belongs is recorded in the upper right corner of each image. Scale bar is 5 nm.

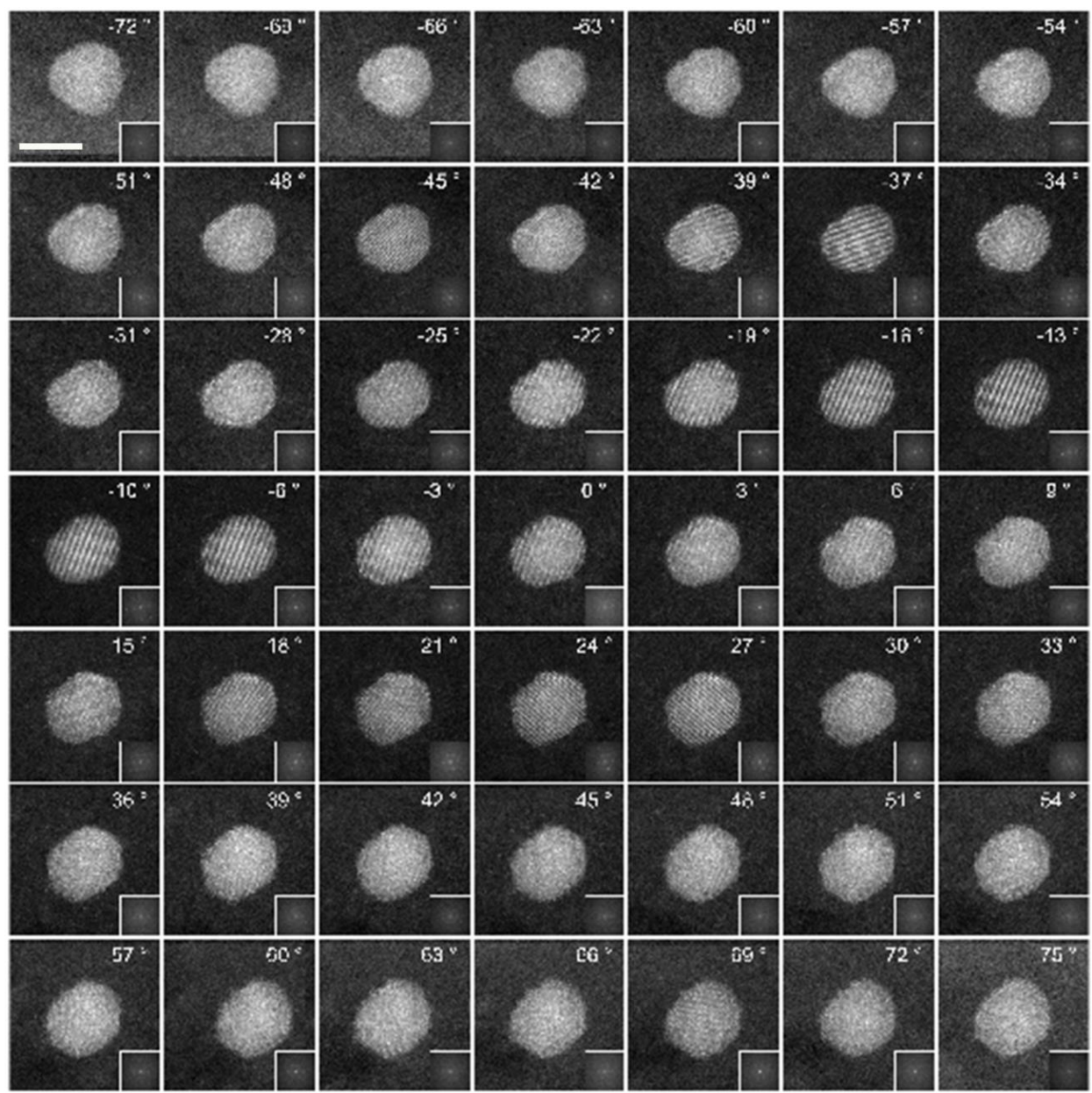

**Fig. S6. Tomographic tilt series of Pd@Pt$_{1\_1.5K}$ NP.**
ADF-STEM images with a tilting range from −72.0° to +75.0°. The value of angle to which the image belongs is recorded in the upper right corner of each image. Scale bar is 5 nm.

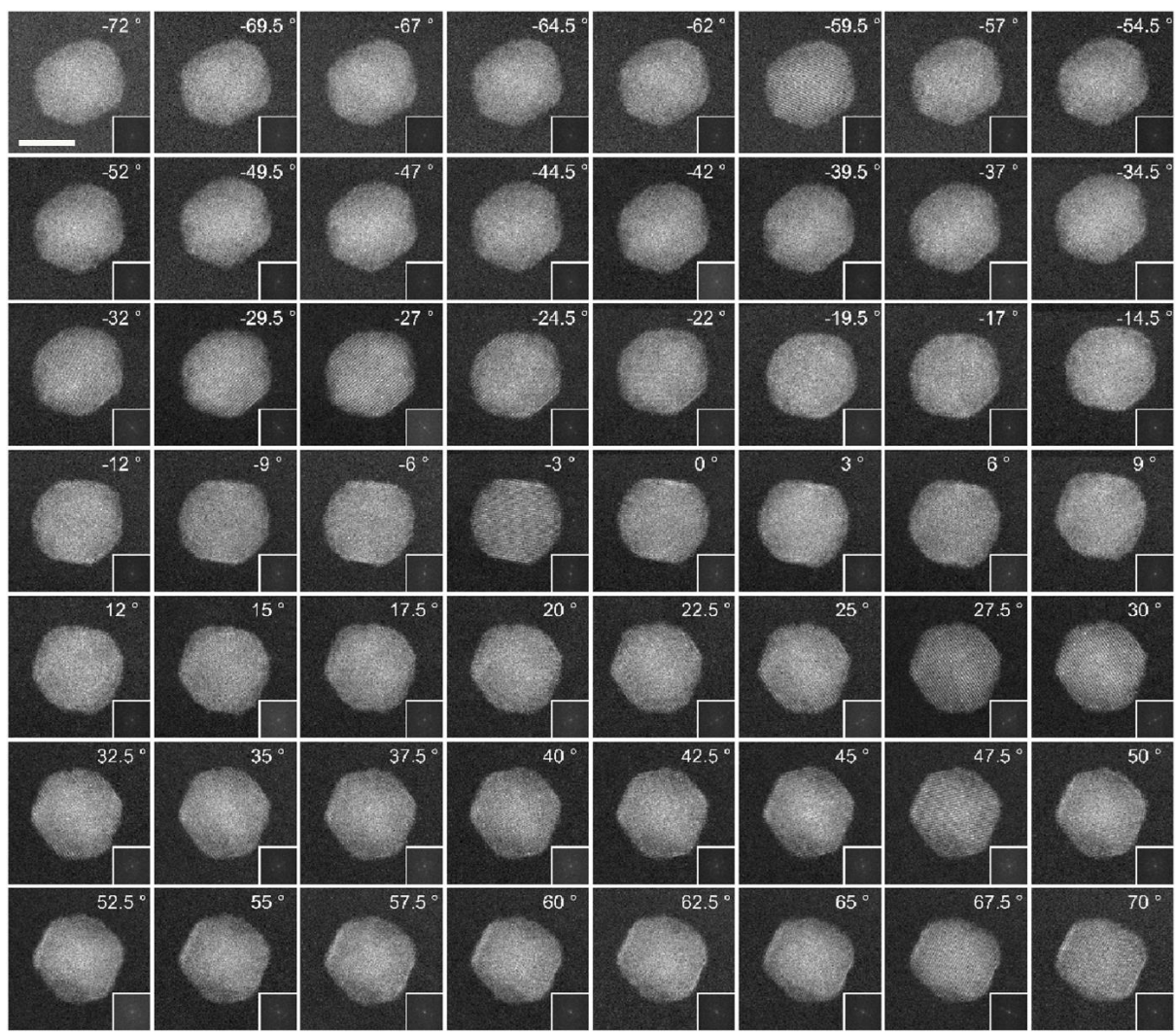

**Fig. S7. Tomographic tilt series of Pd@Pt$_{2\_0}$ NP.**
ADF-STEM images with a tilting range from −72.0° to +70°. The value of angle to which the image belongs is recorded in the upper right corner of each image. Scale bar is 5 nm.

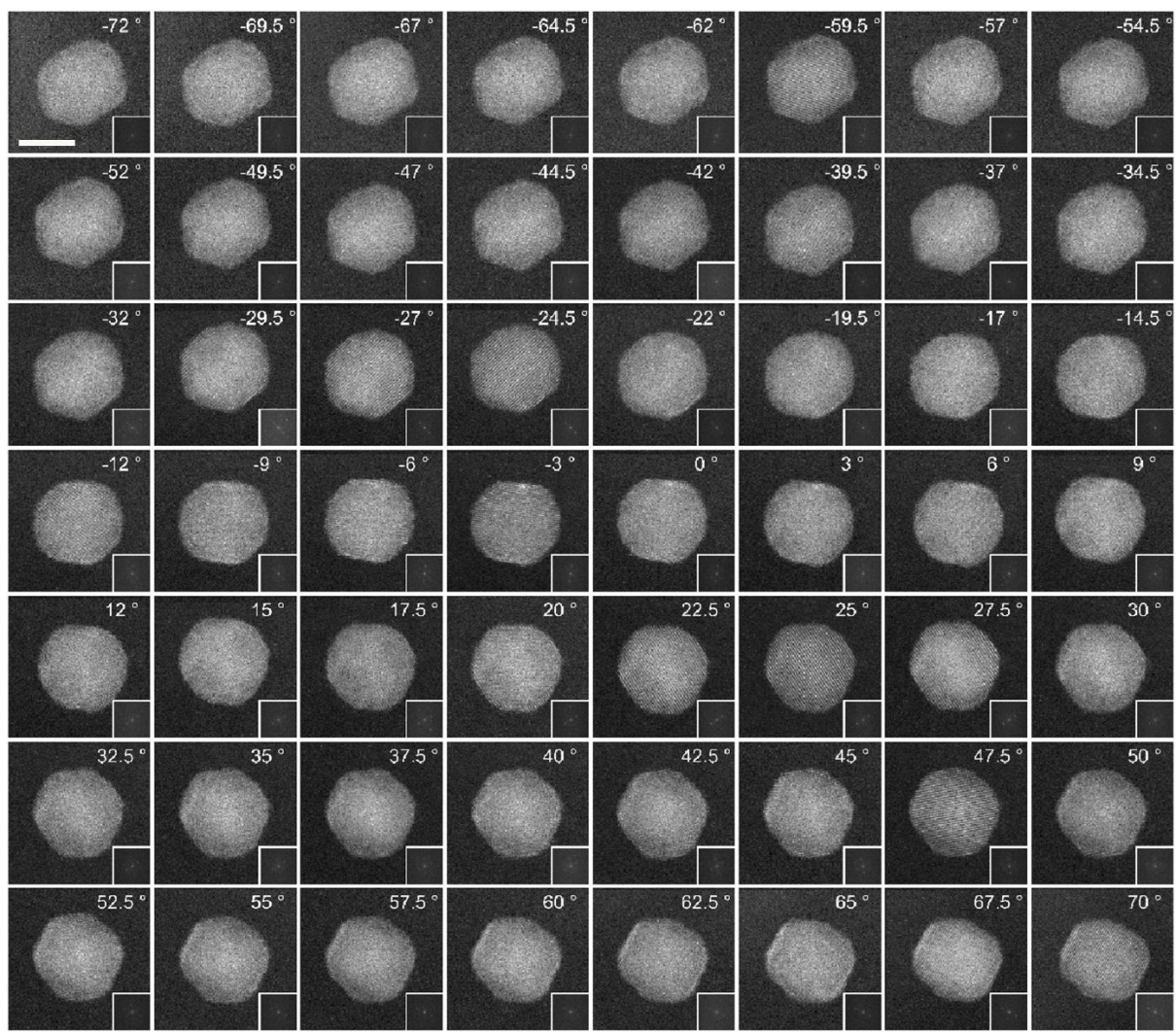

**Fig. S8. Tomographic tilt series of Pd@Pt$_{2\_0.1K}$ NP.**

ADF-STEM images with a tilting range from −72.0° to +70.0°. The value of angle to which the image belongs is recorded in the upper right corner of each image. Scale bar is 5 nm.

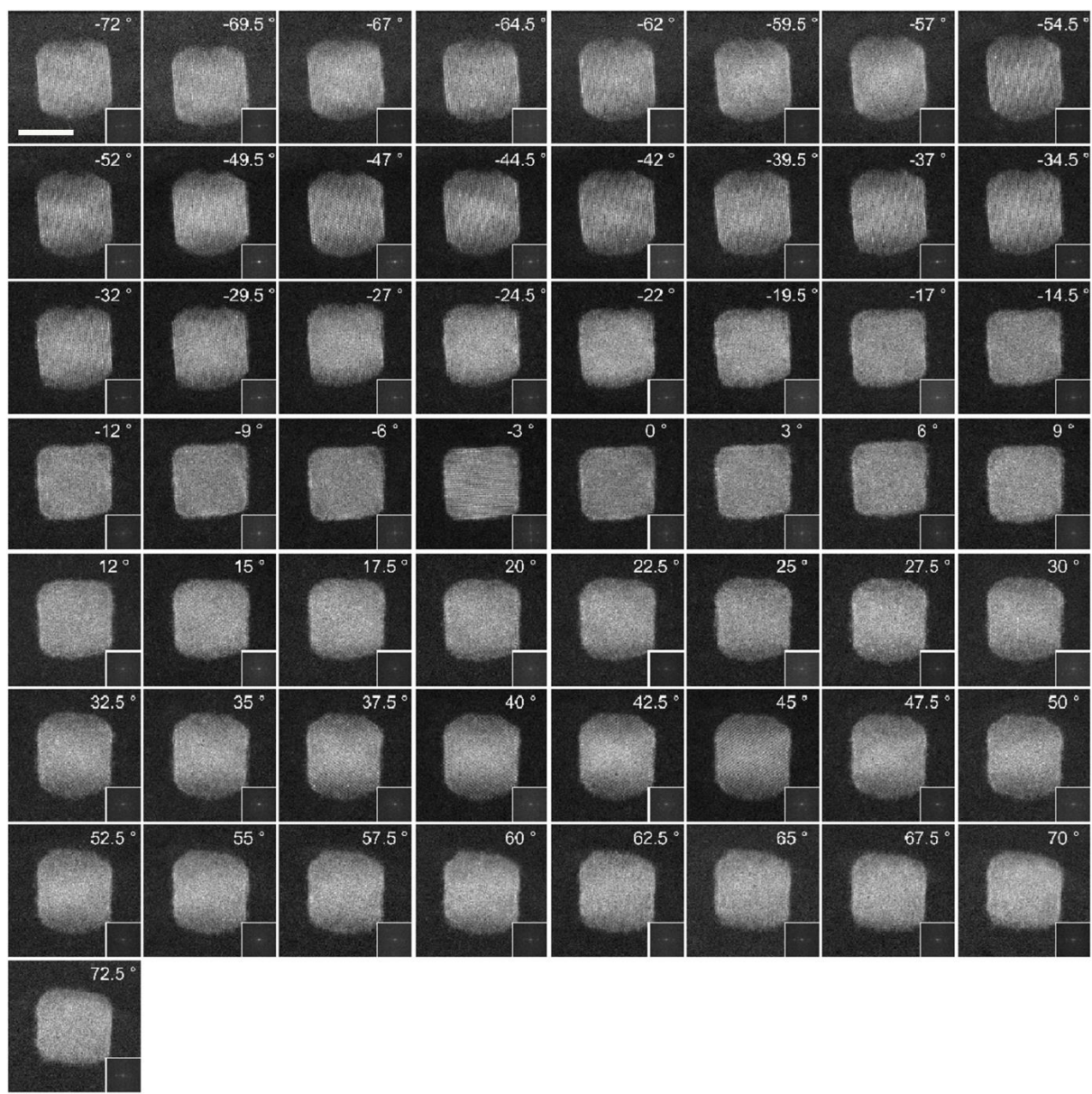

**Fig. S9. Tomographic tilt series of Pd@Pt$_{3\_0}$ NP.**
ADF-STEM images with a tilting range from −72.0° to +72.5°. The value of angle to which the image belongs is recorded in the upper right corner of each image. Scale bar is 5 nm.

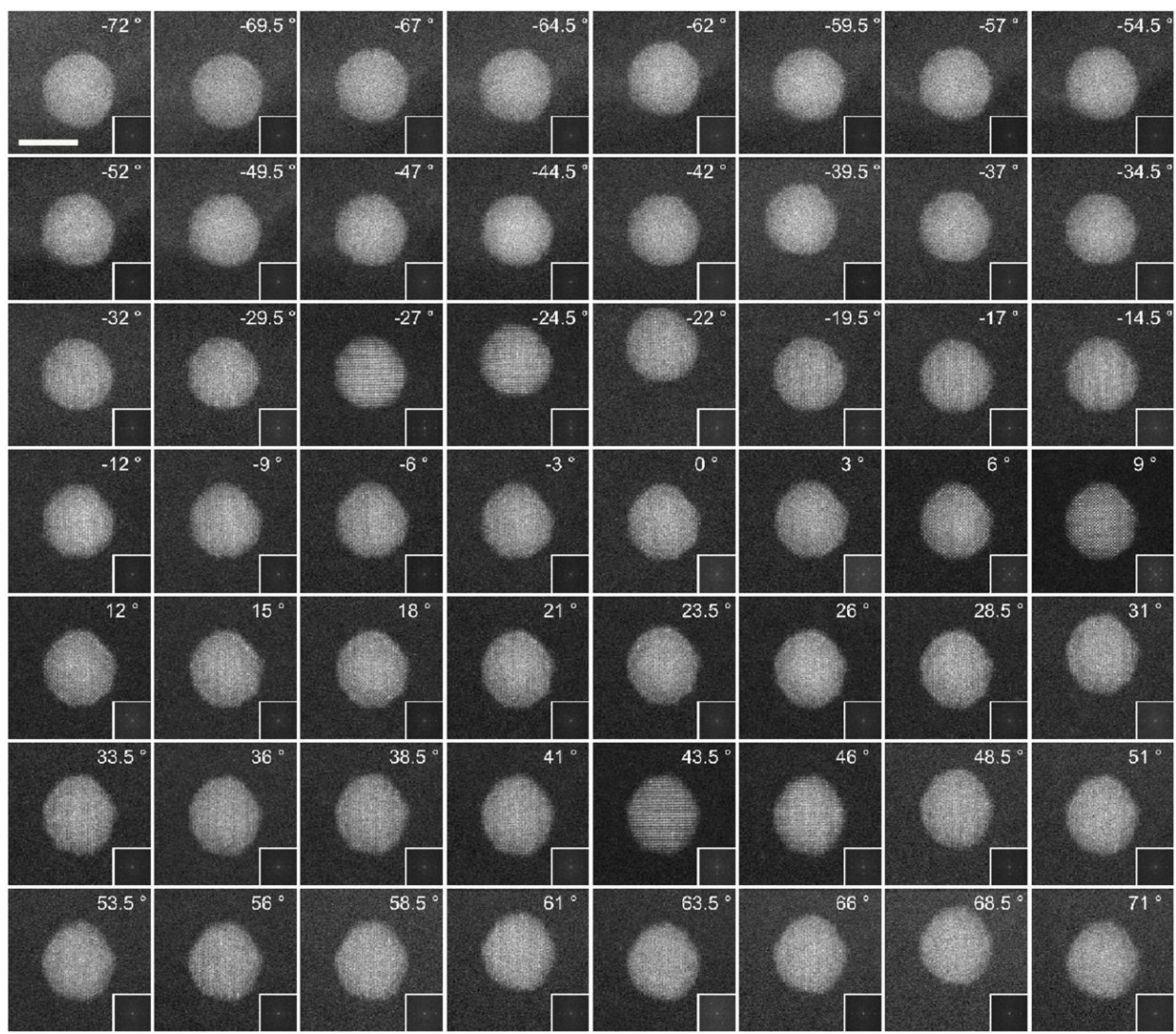

**Fig. S10. Tomographic tilt series of Pd@Pt$_{3\_2.5K}$ NP.**

ADF-STEM images with a tilting range from −72.0° to +71.0°. The value of angle to which the image belongs is recorded in the upper right corner of each image. Scale bar is 5 nm.

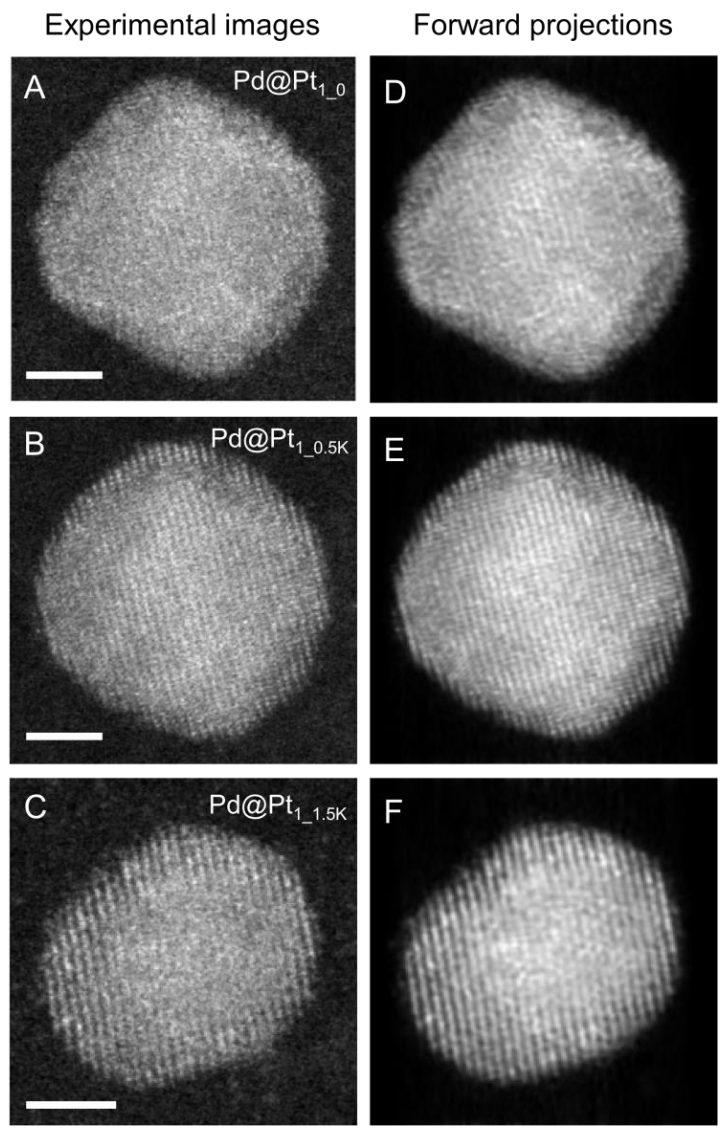

**Fig. S11. Consistency check of particle-1 series.**

(**A** to **C**) ADF-STEM images taken at 0° during tilting experiment for (A) Pd@Pt$_{1\_0}$, (B) Pd@Pt$_{1\_0.5K}$ and (C) Pd@Pt$_{1\_1.5K}$, respectively. (**D** to **F**) Simulated forward projections of the final 3D atomic model at the tilt angles for (D) Pd@Pt$_{1\_0}$, (E) Pd@Pt$_{1\_0.5K}$ and (F) Pd@Pt$_{1\_1.5K}$, respectively. Scale bars for (A) to (F) are 2 nm.

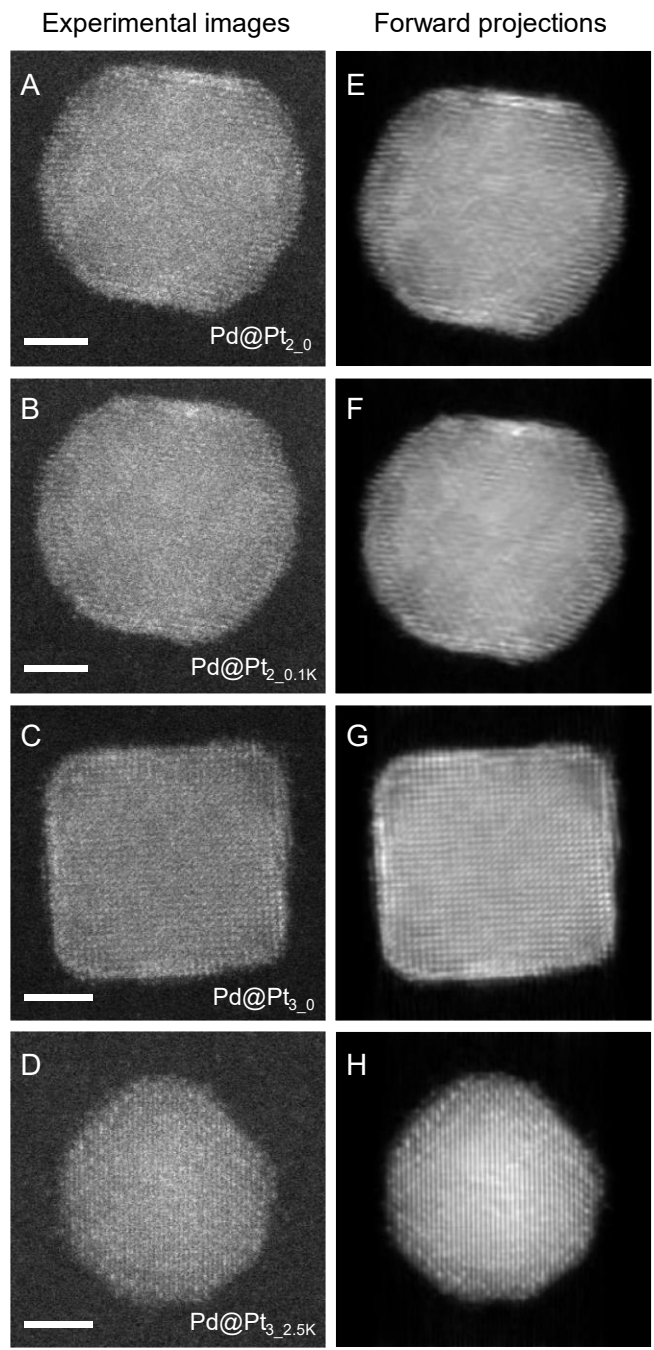

**Fig. S12. Consistency check of particle-2 and -3 series.**

(**A** to **C**) ADF-STEM images taken at 0° during tilting experiment for (A) Pd@Pt$_{2\_0}$, (B) Pd@Pt$_{2\_0.1K}$, (C) Pd@Pt$_{3\_0}$ and (D) Pd@Pt$_{2\_2.5K}$, respectively. (**D** to **F**) Simulated forward projections of the final 3D atomic model at the tilt angles for (E) Pd@Pt$_{2\_0}$, (F) Pd@Pt$_{2\_0.1K}$, (G) Pd@Pt$_{3\_0}$ and (H) Pd@Pt$_{3\_2.5K}$, respectively. Scale bars for (A) to (H) are 2 nm.

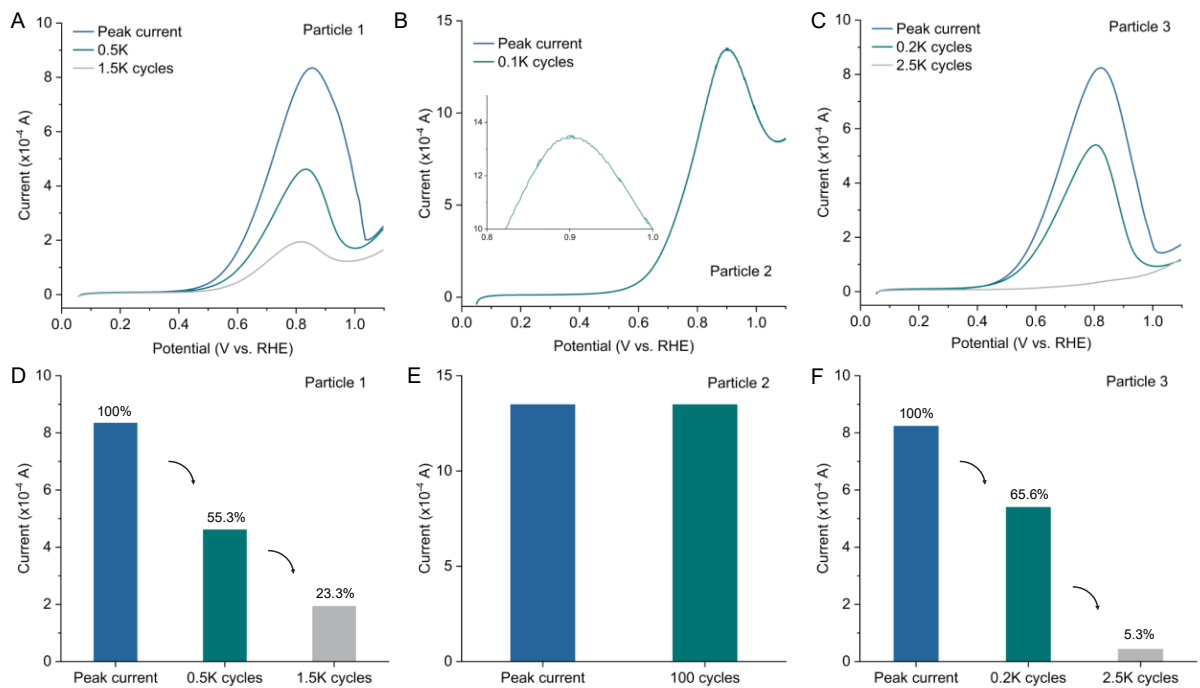

**Fig. S13. Electrocatalytic results for Pd@Pt/CNT nanocatalysts at different periods of ethanol oxidation reaction.**

(**A** to **C**) CV curves of (A) particle-1, (B) particle-2 and (C) particle-3. (**D** to **F**) Current values of (D) particle-1, (E) particle-2 and (F) particle-3. For particle-1, the current decayed from 0.84 mA (peak current) to 0.46 mA (after 500 cycles) and to 0.20 mA (after 1500 cycles). Particle-2 is recorded right after the activation, the current after 100 cycles is equal to the peak current (1.35 mA). For particle-3, the current decayed from 0.82 mA (peak current) to 0.54 mA (after 200 cycles) and to 0.04 mA (after 2,500 cycles).

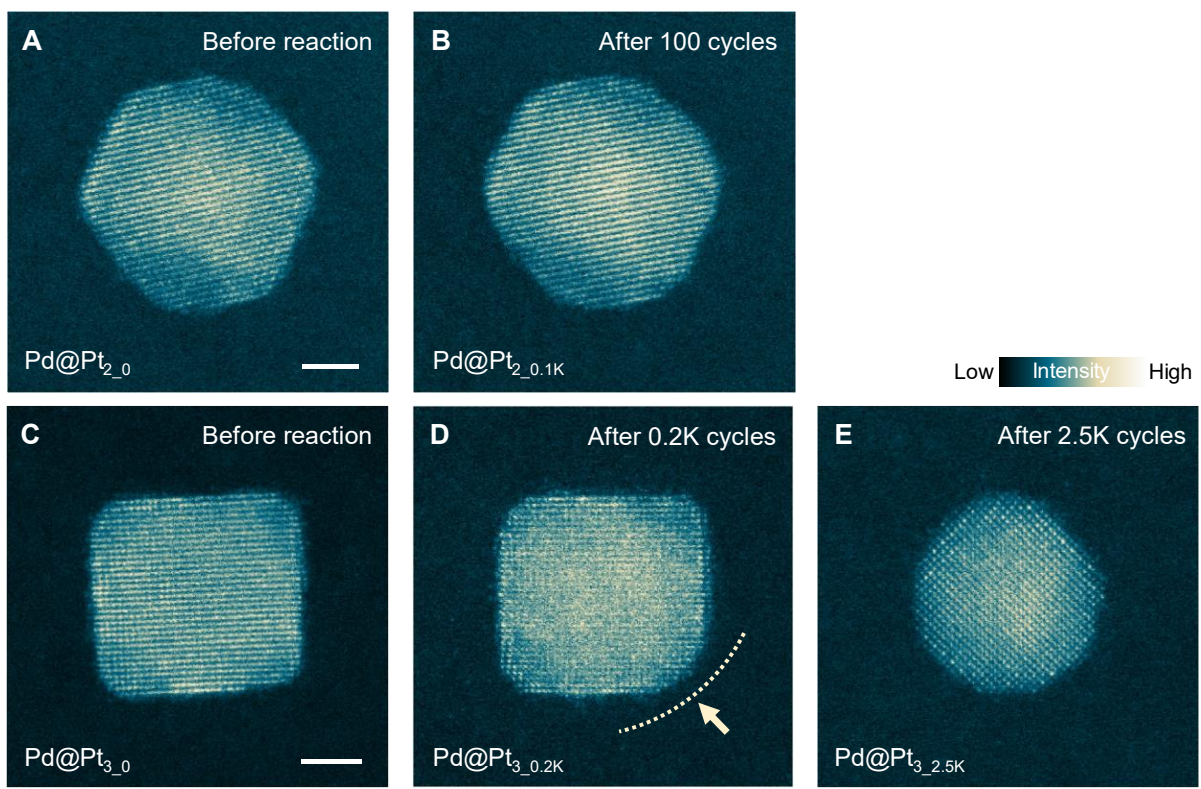

**Fig. S14. Representative images at similar zone axis for particles 2-3.**
(**A** and **B**) ADF-STEM images from tomographic tilt-series of particle-2 (A) before reaction and (B) after 100 cycles. The tilting angles for both of the images are 47.5°. (**A** to **C**) ADF-STEM images from tomographic tilt-series of particle-3 (A) before reaction, (B) after 200 cycles and (C) after 2,500 cycles. The tilting angles for the images are (A) -3°, (B) 6° and (C) -9°, respectively. Images were registered from three frames and normalized. The colormap for (A to E) is placed above (E). Scale bars in are 2 nm. As particle-3 after 200 cycles shows a rounded shape, similar to the result of Pd@Pt$_{1\_0.5K}$ and Pd@Pt$_{2\_0.1K}$, we did not analyze the 3D structure but pushed on the reaction to 2.5K cycles in total to further investigate the 3D structural evolution in longer cycles.

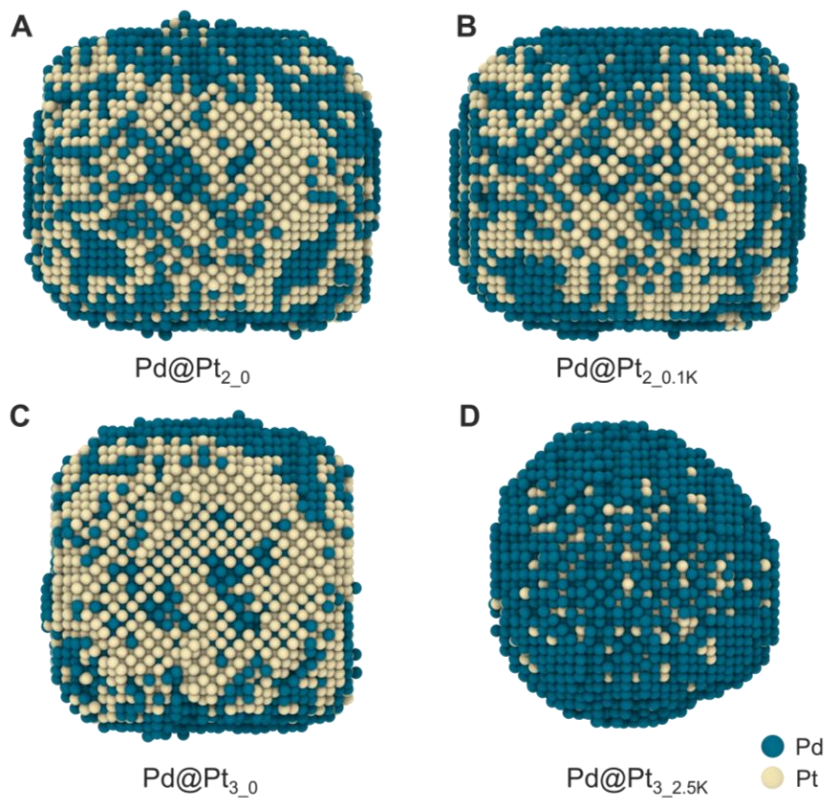

**Fig. S15. The final 3D models for particles 2-3.**

(**A** and **B**) Experimentally derived 3D atomic renderings for particle-2: (A) Pd@Pt$_{2\_0}$ and (B) Pd@Pt$_{2\_0.1K}$. A certain degree of surface reconstruction was observed at the near-surface region. (**C** and **D**) Experimentally derived 3D atomic renderings for particle-3: (C) Pd@Pt$_{3\_0}$ and (D) Pd@Pt$_{3\_2.5K}$. Pd and Pt atoms are depicted as green and yellow spheres respectively. The structures change a lot before and after 2.5K cycles, severe atom leaching and Pd segregation were observed.

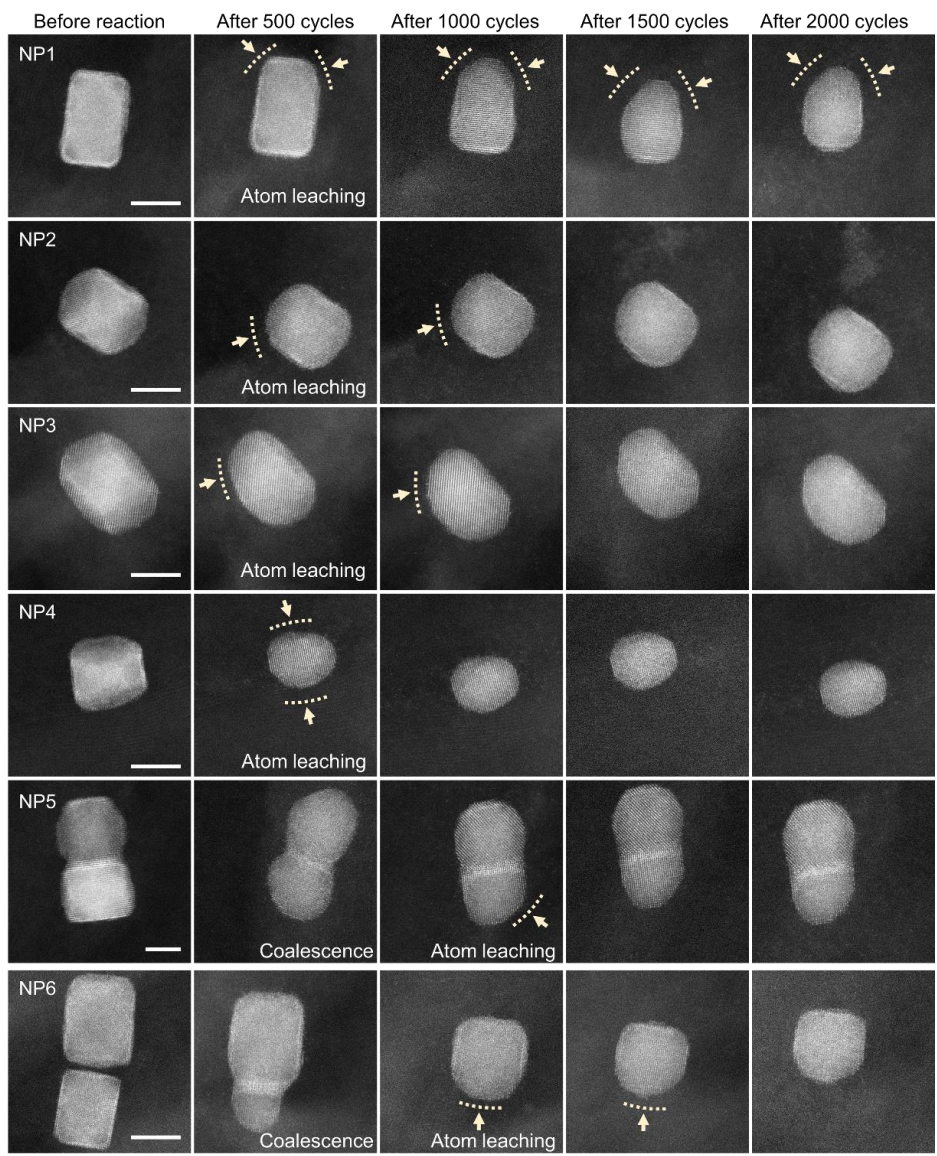

**Fig. S16. Ex situ observations of six batches of NPs before and after 500-2000 cycles of reaction.**

From up to bottom, each row represents one batch of NP, named as NP1, 2, 3, 4, 5 and 6, respectively. From left to right, each of the NP was imaged before reaction and after 500, 1000, 1500 and 2000 cycles, respectively. For NP1-4, obvious structural changes were observed at continuous cycles. For NP5 and NP6, coalescences of two particles close to each other were observed after 500 cycles. The structures still experienced atom leaching after 1000-2000 cycles. The changes are marked by arrows and dashed curves.

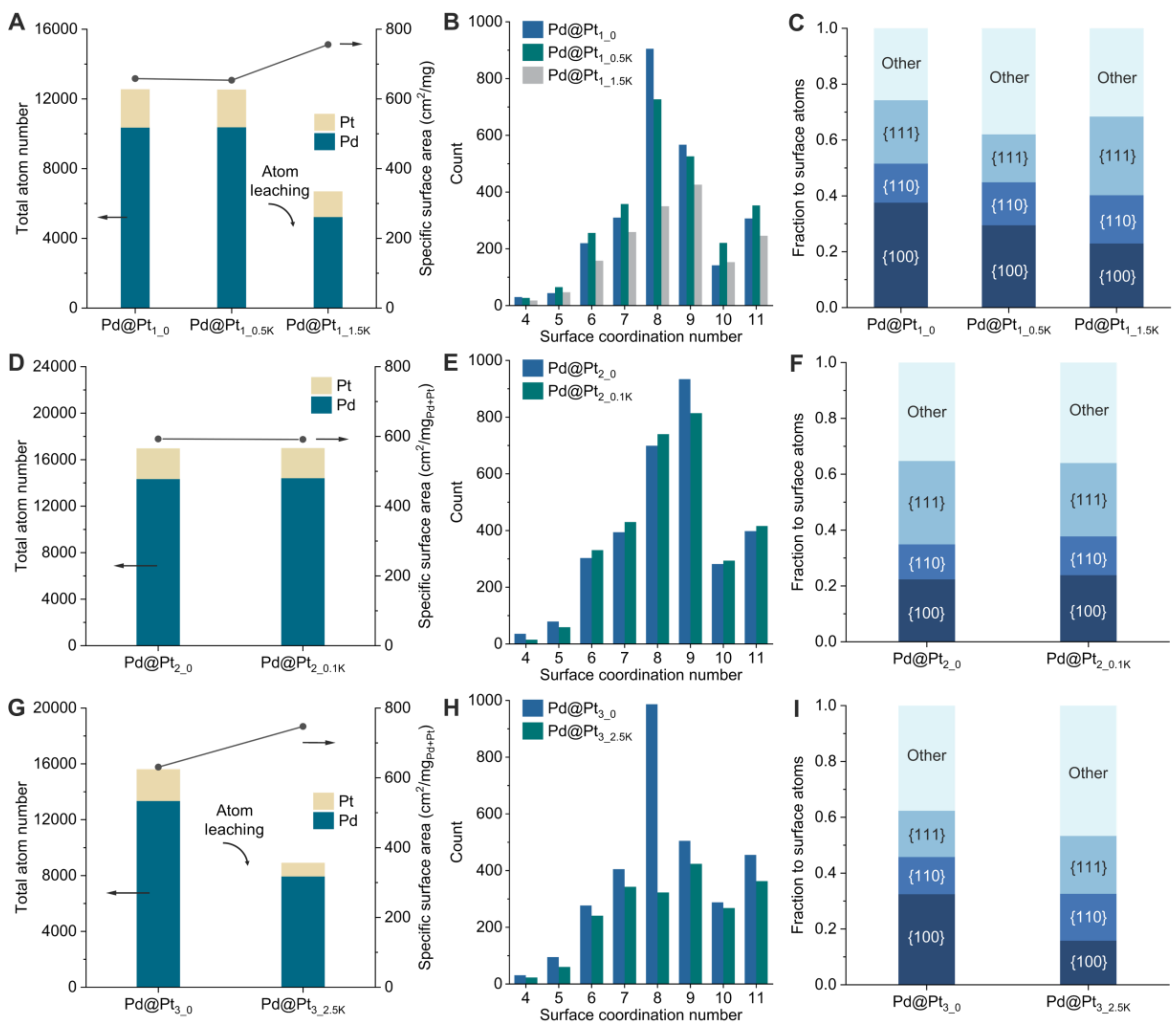

**Fig. S17. Basic structural analysis of Pd@Pt particle 1-3 series.**
(**A** to **C**) Information of (A) total atom numbers and specific surface areas, (B) surface coordination number and (C) surface facet distribution of particle-1. (**D** to **F**) Information of (D) total atom numbers and specific surface areas, (E) surface coordination number and (F) surface facet distribution of particle-2. (**G** to **I**) Information of (G) total atom numbers and specific surface areas, (H) surface coordination number and (I) surface facet distribution of particle-3.

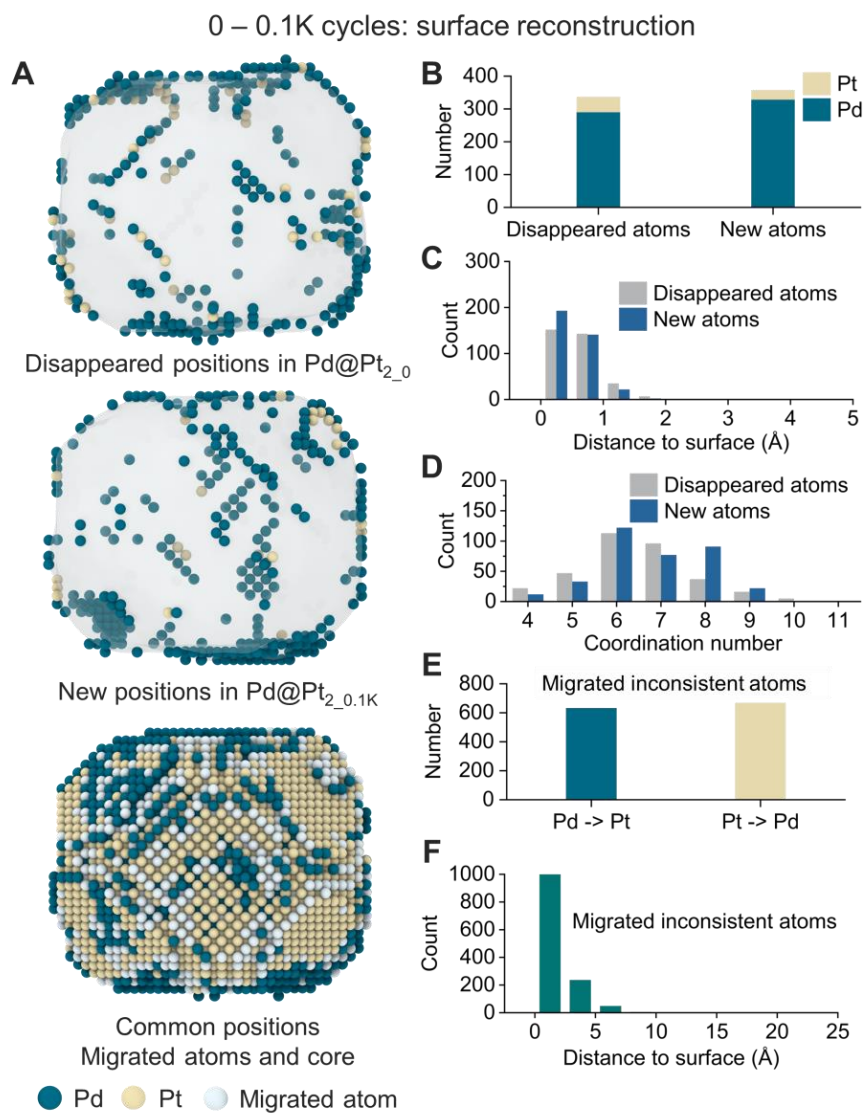

**Fig. S18. Evolutions of positions and chemical species for particle-2.**

(**A**) 3D renderings before and after 100 cycles: disappeared positions (upper panel, positions in Pd@Pt$_{2\_0}$ only), new positions (central panel, only in Pd@Pt$_{2\_0.1K}$), and common positions (lower panel). Green, yellow, and white spheres are denoted as Pd, Pt, and migrated inconsistent atoms at the common positions, respectively. (**B** to **D**) Basic structural analysis of unpaired atoms before after 100 cycles: (B) numbers of atoms, (C) distribution of distances to surface and (D) distribution of coordination numbers. In the calculation of distance to surface: distances of disappeared atoms are calculated by Pd@Pt$_{2\_0}$ surface, and distances of new atoms are calculated by Pd@Pt$_{2\_0.1K}$ surface. (**E** and **F**) Migrated inconsistent atoms at common positions before and after 100 cycles: (E) numbers of atoms and (F) distribution of distances to surface (calculated by Pd@Pt$_{2\_0}$ surface).

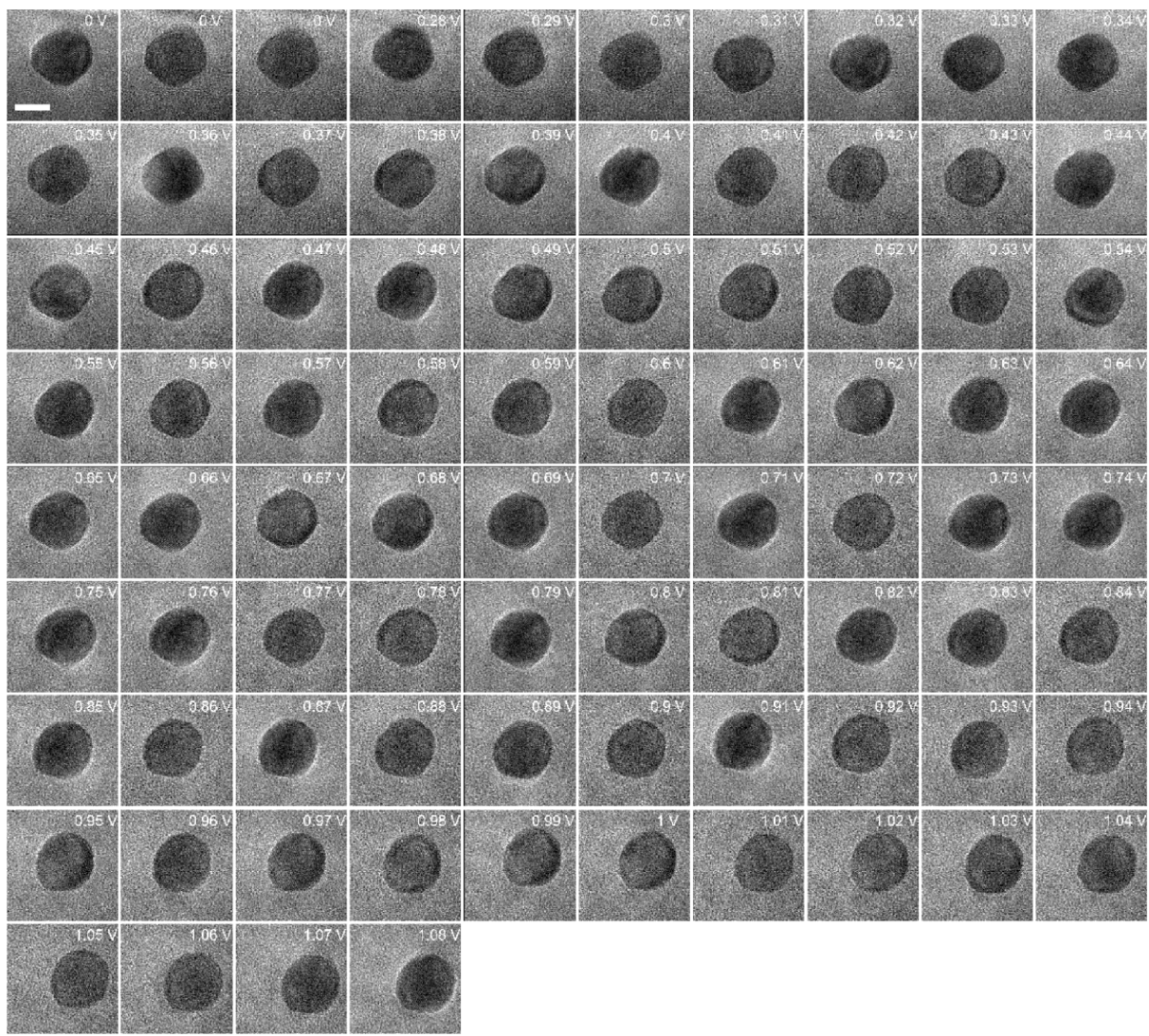

**Fig. S19. Voltage-sequential in situ TEM images of morphological evolution under CV condition.**

TEM images before and during CV conditions from 0.28 V$_{RHE}$ to 1.08 V$_{RHE}$. Scale bar for all panels is 5 nm.

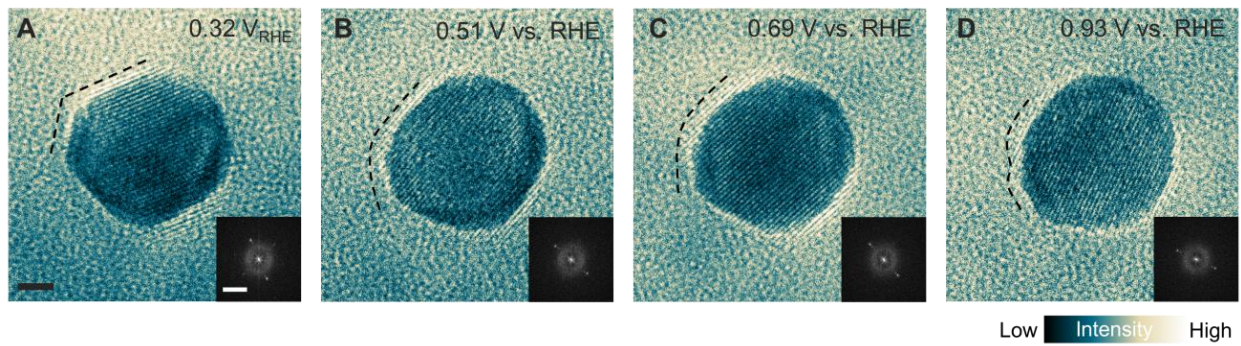

**Fig. S20. In situ electrochemical TEM observations.**

Voltage-sequential in situ TEM images of morphological evolution under cyclic voltammetry from (A) 0.32 $V_{RHE}$, (B) 0.51 $V_{RHE}$, (C) 0.69 $V_{RHE}$ to (D) 0.93 $V_{RHE}$. Corresponding diffraction patterns are shown in insets. Scale bars for all panels in (A to D) are 2 nm for the TEM images and 5 $nm^{-1}$ for the FFT patterns. Colormap for (A) to (D) is placed below (D).

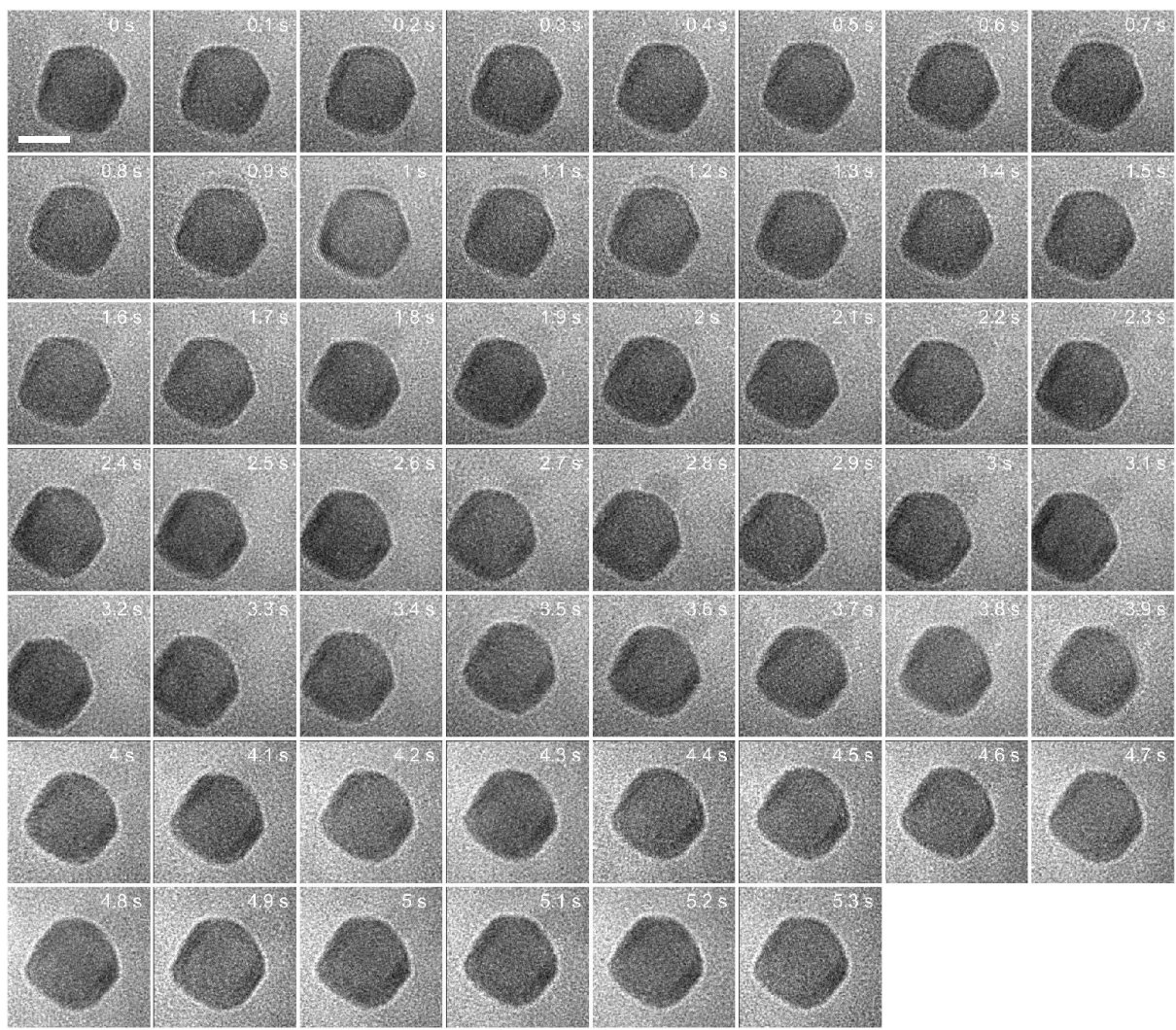

**Fig. S21. Time-sequential in situ TEM images of morphological evolution under chronoamperometry (CA) condition at 0.68 V$_{RHE}$.**

We kept the interested region (upper right region of the NP) always in the field of view, while part of the particle drifted out of the field of view. Scale bar for all panels is 5 nm.

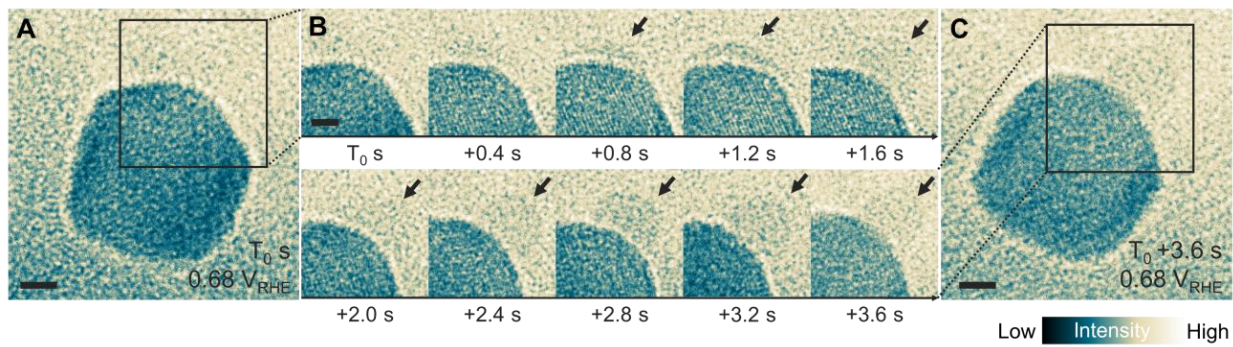

**Fig. 22. In situ electrochemical TEM observations.**

Time-sequential in situ TEM images under a voltage of 0.68 $V_{RHE}$ (A) before and (B and C) during atom leaching happens. Arrows point to the locations of fluctuating clusters. Colormap for (A) to (C) is placed below (C). Scale bars are 2 nm for (A) to (C).

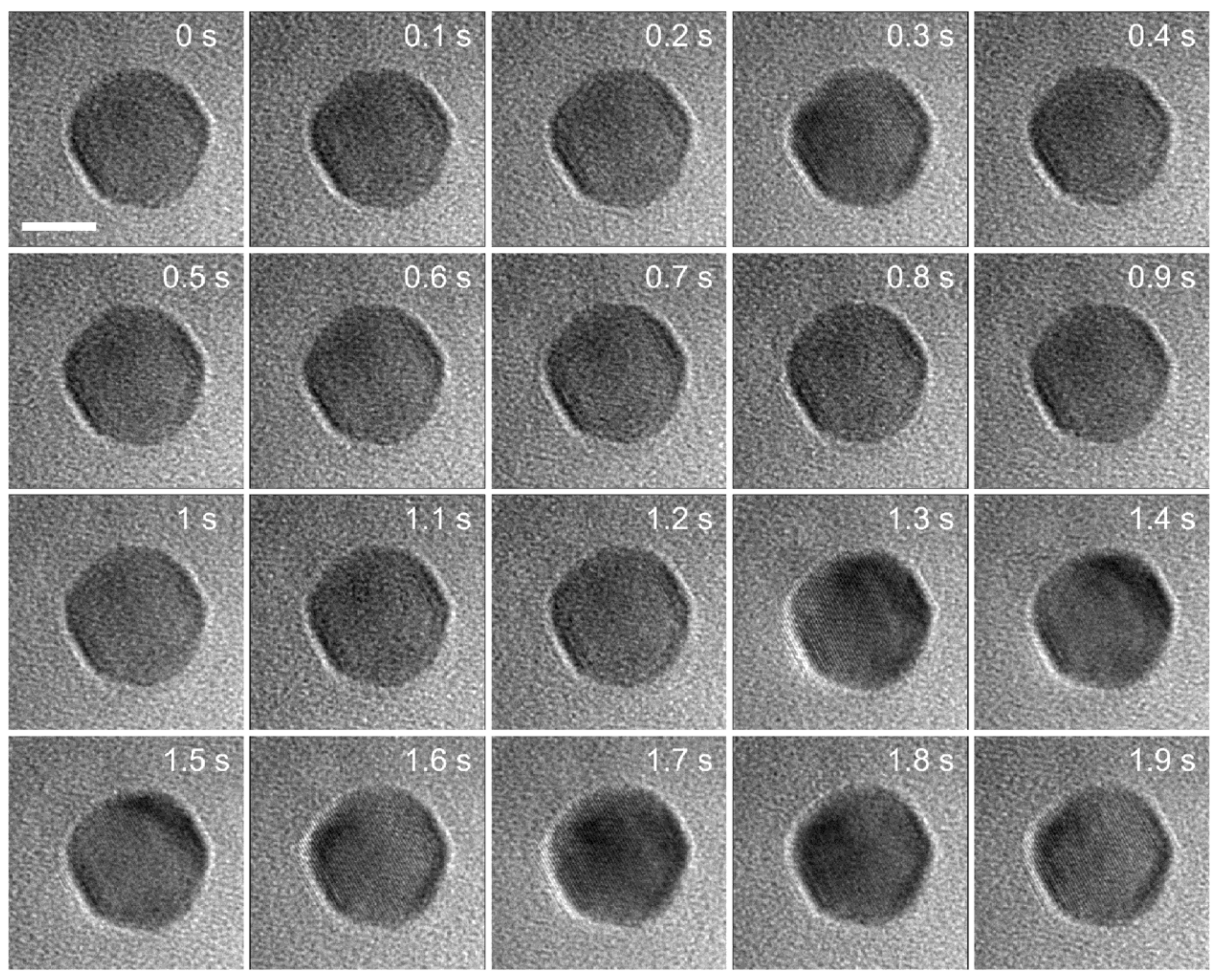

**Fig. S23. Time-sequential in situ TEM images of morphological evolution under CA condition at 0.68 V$_{RHE}$.**
Scale bar for all panels is 5 nm.

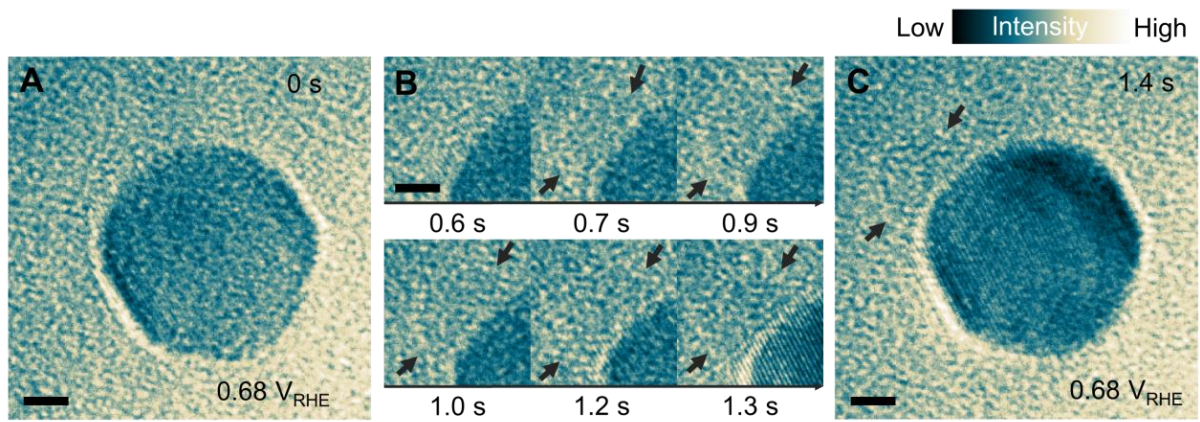

**Fig. S24. Time-sequential in situ TEM images under a voltage of 0.68 V$_{RHE}$.**
TEM images (A) before and (B and C) during atom leaching happens. Arrows point to the locations of fluctuating clusters. Colormap for (A) to (C) is placed above (C). Scale bars are 2 nm for (A) to (C).

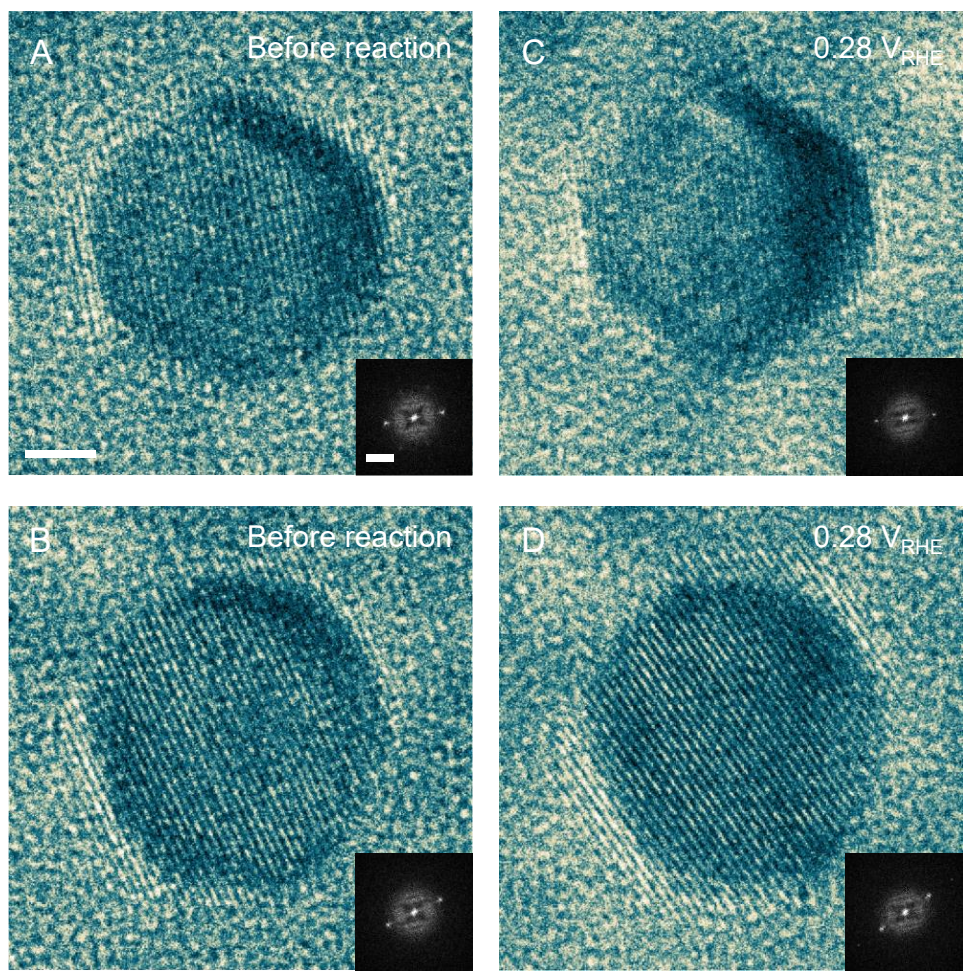

**Fig. S25. In situ electrochemical TEM observations for two NPs before and under CA test with a lower potential of 0.28 V$_{RHE}$.**

TEM images (A and B) before and (C and D) during CA test. Scale bar for all panels in (A) to (D) is 2 nm.

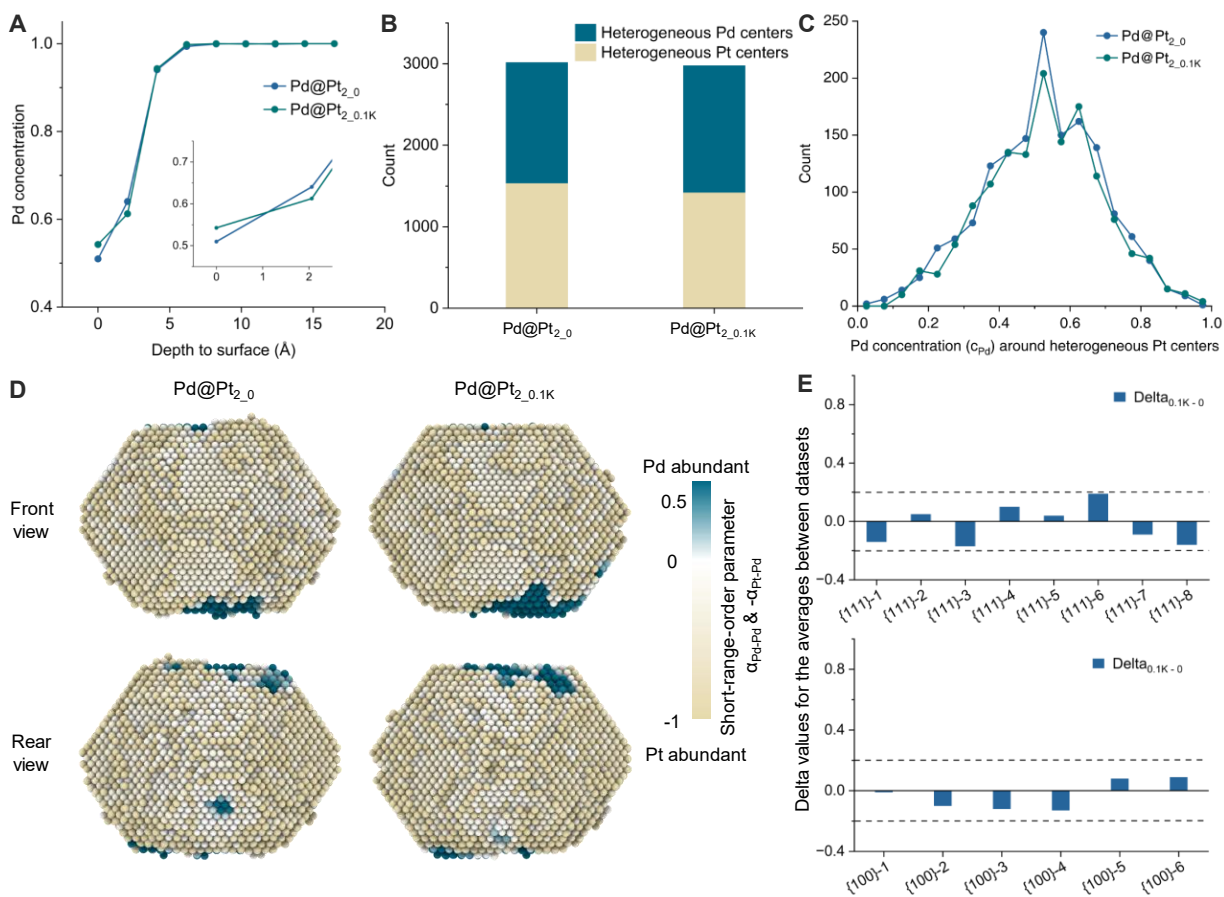

**Fig. S26. Evolutions of local chemical heterogeneity and CSROP for particle-2.**
(**A**) Shell-by-shell Pd concentrations throughout the evolutions. (**B**) Number of heterogeneous centers, where a center is defined as one atom coordinated by both Pd and Pt neighbors. (**C**) Overall distribution of local Pd concentrations ($c_{Pd}$) around heterogeneous Pt centers. $c_{Pd}$ (x axis) represents Pd/(Pd+Pt) atomic fractions in coordination shells, indicating local Pd occurrence probabilities around central atoms. (**D**) Spatial renderings of CSROPs (front/rear views across three cycles). Green represents the Pd-abundant regions, yellow represents the Pt-abundant regions and white represents the regions equal to the averaged Pd concentration in the system. Values are normed to 0.5 or -1, respectively. (**E**) Delta values for the orientation-classified averages of CSROPs. Surface atoms were classified into eight {111} orientations (top) and six {100} orientations (bottom), with values averaged within each region (figure S28). Then the delta values of $Delta_{0.1K-0}$ are obtained by calculating the differences of averages between Pd@Pt$_{2\_0.1K}$ and Pd@Pt$_{2\_0}$.

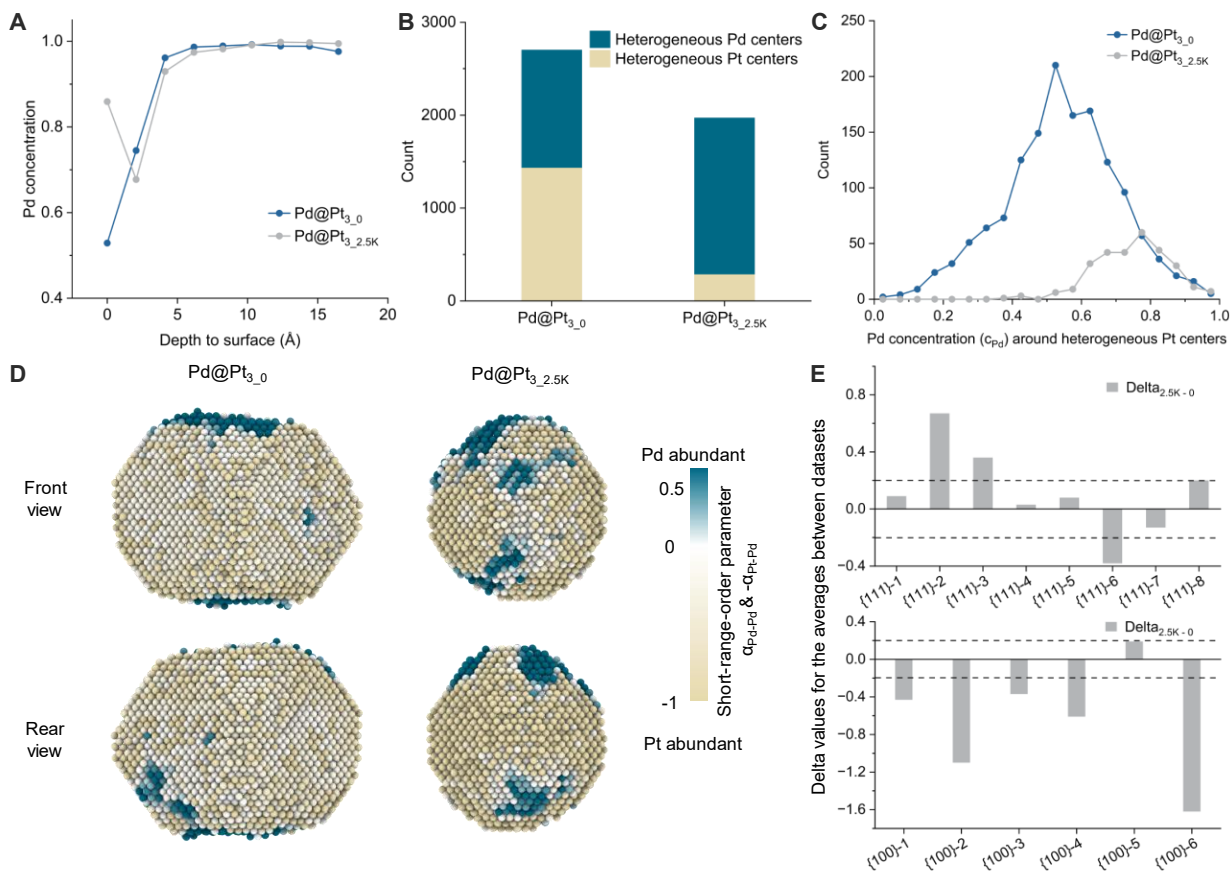

**Fig. S27. Evolutions of chemical heterogeneity and CSROP for particle-3.**
(**A**) Shell-by-shell Pd concentrations throughout the evolutions. (**B**) Number of heterogeneous centers, where a center is defined as one atom coordinated by both Pd and Pt neighbors. (**C**) Overall distribution of local Pd concentrations ($c_{Pd}$) around heterogeneous Pt centers. $c_{Pd}$ (x axis) represents Pd/(Pd+Pt) atomic fractions in coordination shells, indicating local Pd occurrence probabilities around central atoms. (**D**) Spatial renderings of CSROPs (front/rear views across three cycles). Green represents the Pd-abundant regions, yellow represents the Pt-abundant regions and white represents the regions equal to the averaged Pd concentration in the system. Values are normed to 0.5 or -1, respectively. (**E**) Delta values for the orientation-classified averages of CSROPs. Surface atoms were classified into eight {111} orientations (top) and six {100} orientations (bottom), with values averaged within each region (figure S28). Then the delta values of Delta$_{0.1K-0}$ are obtained by calculating the differences of averages between Pd@Pt$_{3\_2.5K}$ and Pd@Pt$_{3\_0}$.

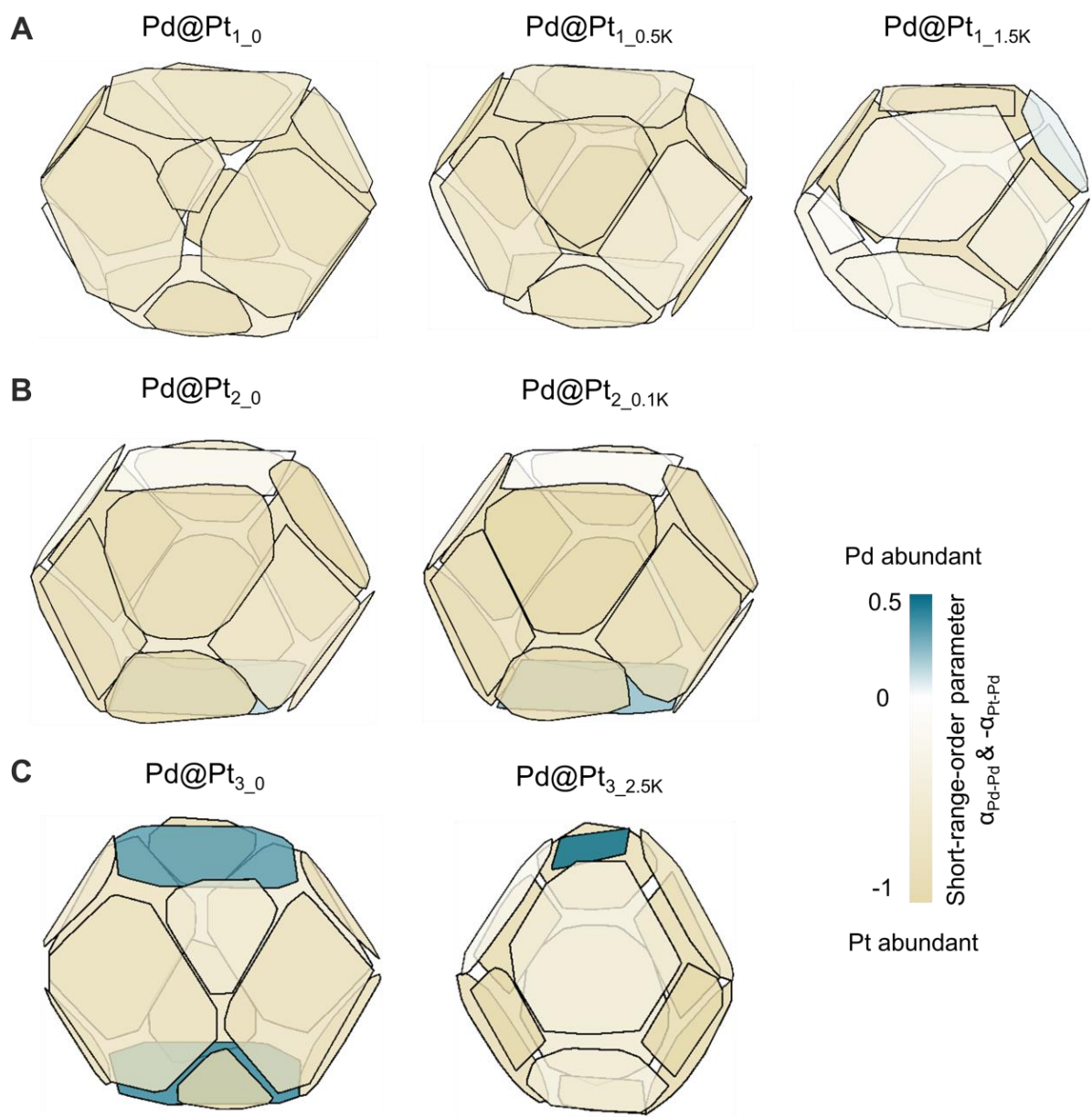

**Fig. S28. Renderings of averaged CSROPs in different orientations.**
(**A** to **C**) The surface atoms were classified into 8 regions of {111} orientations (upper panel) and 6 regions of {100} orientations (lower panel). Then the CSROPs of atoms in different regions were averaged for (A) particle-1, (B) particle-2 and (C) particle-3, respectively. Green colors represent the Pd-abundant regions, yellow colors represent the Pt-abundant regions, and white colors represent the regions close to the averaged Pd concentration in the system. Values are normed to 0.5 or 1, respectively.

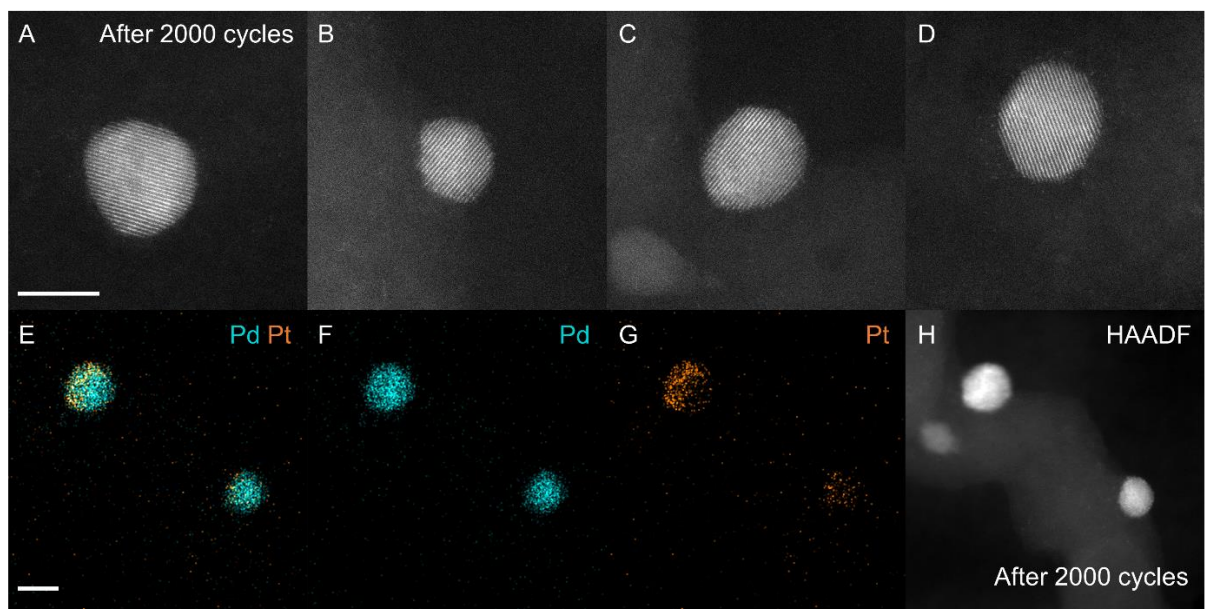

**Fig. S29. Characterization of Pd@Pt/CNT nanocatalysts after 2,000 cycles.**
(**A** to **D**) ADF-STEM images of Pd@Pt after 2,000 cycles. Scale bars are 5 nm for (A) to (D). (**E** to **H**) EDS mapping of Pd@Pt/CNT nanocatalysts after 2,000 cycles: images of (E) merged Pd and Pt signals, (F) Pd, (G) Pt and (H) ADF signals. Scale bars are 5 nm for (E) to (H). The core-shell features are lost and the sizes of particles are smaller after long cycles. Pt signals of NP in (E) and (G) show asymmetric features, which indicates the exist of element redistribution after long cycles.

**Table S1. Volumes of inks used in sample preparation**

| Name of particle series | Particle-1 | Particle-2 | Particle-3 |
|---|---|---|---|
| Pd@Pt/CNT ink (μL) | 1 | 2 | 1 |

**Table S2. AET data and analysis**

| Particle name | Particle-1 | | | Particle-2 | | Particle-3 | |
|---|---|---|---|---|---|---|---|
| | $NP_{1\_0}$ | $NP_{1\_0.5K}$ | $NP_{1\_1.5K}$ | $NP_{2\_0}$ | $NP_{2\_0.1K}$ | $NP_{3\_0}$ | $NP_{3\_2.5K}$ |
| **Data collection*** | | | | | | | |
| Tilt range (°) | -72° / 72.5° | -69° / 75.5° | -72° / 75° | -72° / 70° | -72° / -70° | -72° / 72.5° | -72° / 71° |
| Electron dose ($10^5$ e$^-$ Å$^{-2}$) | 3.0 | 2.8 | 2.5 | 2.9 | 2.9 | 3.0 | 2.9 |
| **Refinement†** | | | | | | | |
| R (%)‡ | 6.8 | 6.4 | 8.7 | 5.2 | 4.9 | 5.6 | 6.9 |
| **Statistics of atoms** | | | | | | | |
| Number of Pd | 10353 | 10373 | 5222 | 14331 | 14405 | 13339 | 7930 |
| Number of Pt | 2205 | 2163 | 1475 | 2641 | 2588 | 2284 | 985 |
| **Common positions** | | | | | | | |
| Number of positions | 12108 | 12108/6172§ | 6172 | 16634 | 16634 | /# | /# |
| Ratio to total (%) | 96.4 | 96.6/49.2§ | 92.2 | 98.0 | 97.8 | /# | /# |
| **Consistent atoms** | | | | | | | |
| Number of Pd | 9423 | 9423/4569¶ | 4569 | 13407 | 13407 | /# | /# |
| Number of Pt | 1405 | 1405/251¶ | 251 | 1925 | 1925 | /# | /# |
| Ratio to total (%) | 89.4 | 89.4/78.1¶ | 78.1 | 92.2 | 92.2 | /# | /# |

\* Operation parameters of datasets are: 300 kV voltage, 25 mrad convergence semi-angle, 40.6 mrad HAADF detector inner semi-angle and 200 mrad HAADF detector outer semi-angle, the pixel size is 0.352 Å.

† Real space iterative reconstruction (RESIRE) algorithm (*53*) was used to reconstruct each experimental tilt series, with oversampling ratio being 4 and number of iterations being 300.

‡ A R-factor is used to calculate the difference between calculated and measured projections (*53*).

§ 12108 and 96.6% are the common atoms and ratio to total atoms between Pd@Pt$_{1\_0}$ and Pd@Pt$_{1\_0.5K}$; and 6172 and 49.2% are the consistent atoms and ratio to total atoms between Pd@Pt$_{1\_0.5K}$ and Pd@Pt$_{1\_1.5K}$, respectively.

¶ 9423, 1405 and 89.4% are the consistent atoms and ratio to total atoms between Pd@Pt$_{1\_0}$ and Pd@Pt$_{1\_0.5K}$; and 4569, 251 and 78.1% are the consistent atoms and ratio to total atoms between Pd@Pt$_{1\_0.5K}$ and Pd@Pt$_{1\_1.5K}$, respectively.

# For particle-3, due to the drastic changes between Pd@Pt$_{3\_0}$ and Pd@Pt$_{3\_2.5K}$, it is hard to find the common positions of the two structures for further analysis. In the meantime, as particle-3 transformed from a particle with Pt-abundant surface into a smaller one with Pd-abundant surface (fig. S15), the decrease in size and the Pd segregation implies inner Pd atoms undergo a dramatic change at this stage.

**Table S3. The delta values of CSROPs between datasets**

| Orientations | Particle-1 | | Particle-2 | Particle-3 |
|---|---|---|---|---|
| | $Delta_{0.5K-0}$* | $Delta_{1.5K-0.5K}$† | $Delta_{0.1K-0}$‡ | $Delta_{2.5K-0}$§ |
| {100}-1¶ | -0.03 | 0.62# | -0.01 | -0.43 |
| {100}-2 | 0.01 | -0.01 | -0.10 | -1.1 |
| {100}-3 | 0.19 | 0.18 | -0.12 | -0.37 |
| {100}-4 | -0.16 | -0.02 | -0.13 | -0.61 |
| {100}-5 | 0.02 | -0.06 | 0.08 | 0.19 |
| {100}-6 | 0.01 | -0.02 | 0.09 | -1.62 |
| {111}-1 | -0.11 | 0.55 | -0.14 | 0.09 |
| {111}-2 | 0.12 | 0.47 | 0.05 | 0.67 |
| {111}-3 | 0.15 | 0.30 | -0.17 | 0.36 |
| {111}-4 | -0.03 | 0.73 | 0.10 | 0.03 |
| {111}-5 | 0.07 | -0.22# | 0.04 | 0.08 |
| {111}-6 | -0.05 | 0.61 | 0.19 | -0.38 |
| {111}-7 | -0.10 | 0.02 | -0.09 | -0.13 |
| {111}-8 | 0.27 | 0.27 | -0.16 | 0.20 |

\* $Delta_{0.5K-0}$ means the delta values of averaged CSROPs between Pd@Pt$_{1\_0.5K}$ and Pd@Pt$_{1\_0}$.
† $Delta_{1.5K-0.5K}$ means the delta values of averaged CSROPs between Pd@Pt$_{1\_1.5K}$ and Pd@Pt$_{1\_0.5K}$.
‡ $Delta_{0.1K-0}$ means the delta values of averaged CSROPs between Pd@Pt$_{2\_0.1K}$ and Pd@Pt$_{2\_0}$.
§ $Delta_{2.5K-0}$ means the delta values of averaged CSROPs between Pd@Pt$_{3\_2.5K}$ and Pd@Pt$_{3\_0}$.
¶ The assigned number for {100} or {111} orientations shown in Fig. 3E, fig. 19E and fig. 20E
# While the delta values in particle-2 are all smaller than 0.20, indicating little changes during the 100 cycles. The delta values higher than 0.2 or lower than -0.2 in the table are highlighted in red or green, respectively.

**Movie S1**

Raw TEM images in Fig. 4, A to D and fig. S23, under CV condition from 0.28 V$_{RHE}$ to 1.08 V$_{RHE}$. The arrow in the movie points to the rounded places.

**Movie S2.**

Raw TEM images in Fig. 4, E to G and fig. S24, under CA condition at 0.68 V$_{RHE}$. The arrow in the movie points to the locations of fluctuating clusters.

**Movie S3.**

Raw TEM images in figs. S25 and S26, under CA condition at 0.68 V$_{RHE}$. The arrow in the movie points to the locations of fluctuating clusters.

**Movie S4.**

Raw TEM images in fig. S27 A and B, with no bias.

**Movie S5.**

Raw TEM images in fig. S27 A and B, under CA condition at 0.28 V$_{RHE}$.